\documentclass[a4paper,fleqn]{cas-dc}

\usepackage[authoryear,longnamesfirst]{natbib}
\usepackage{graphicx}
\usepackage{subcaption}
\usepackage{soul} 
\usepackage{color, xcolor} 
\def\tsc#1{\csdef{#1}{\textsc{\lowercase{#1}}\xspace}}
\tsc{electric vehicle}

\begin{document}
\let\WriteBookmarks\relax
\def\floatpagepagefraction{1}
\def\textpagefraction{.001}

\shorttitle{}

\shortauthors{}

\title [mode = title]{The impact of large-scale EV charging on the real-time operation of distribution systems: A comprehensive review}                      
\tnotemark[1,2]


% \tnotetext[2]{}

\author[1]{Zhe Yu}[type=editor,
                        auid=000,bioid=1,
                        prefix= ,
                        role= ,
                        orcid=0009-0008-3518-9057]

%\cormark[1]


\ead{zhe-ee.yu@connect.polyu.hk}

\ead[url]{ }

\credit{Conceptualization, Methodology, Writing, Visualization}

\affiliation[1]{organization={Department of Electrical and Electronic Engineering},
    addressline={The Hong Kong Polytechnic University}, 
    city={Kowloon},
    % citysep={}, % Uncomment if no comma needed between city and postcode
    postcode={999077}, 
    country={Hong Kong Special Administrative Region}}

\author[1]{Chuang Yang}[orcid=0000-0002-4086-8387]
\ead{c1yang@polyu.edu.hk}
\ead[URL]{}
\credit{Reviewing}
%\author[2,3]{ }[%
   %role= ,
   %suffix= ,
%\ead{}
%\ead[URL]{}

\author[1]{Qin Wang}
\cormark[1]
%\fnmark[1,3]
\ead{qin-ee.wang@polyu.edu.hk}
\ead[URL]{}
\credit{Supervision, Conceptualization, Methodology, Reviewing, Visualization}

\cortext[cor1]{Corresponding author}
%\cortext[cor2]{Principal corresponding author}

% \fntext[fn1]{}
% \fntext[fn2]{}

\nonumnote{}

\begin{abstract}
With the large-scale integration of electric vehicles (EVs) in the distribution grid, the unpredictable nature of EV charging introduces considerable uncertainties to the grid's real-time operations. This can exacerbate load fluctuations, compromise power quality, and pose risks to the grid's stability and security. However, due to their dual role as controllable loads and energy storage devices, EVs have the potential to mitigate these fluctuations, balance the variability of renewable energy sources, and provide ancillary services that support grid stability. By leveraging the bidirectional flow of information and energy in smart grids, the adverse effects of EV charging can be minimized and even converted into beneficial outcomes through effective real-time management strategies. This paper explores the negative impacts of EV charging on the distribution system's real-time operations and outlines methods to transform these challenges into positive contributions. Additionally, it provides an in-depth analysis of the real-time EV management system, focusing on state estimation, system frameworks, performance improvement and emerging technologies.
\end{abstract}

% \includegraphics{figs/grabs.pdf}

\begin{highlights}
\item A detailed analysis of EV real-time management system is presented. 
\item The grid management strategies to reverse negative impacts from EV integration to positive outcomes are discussed.
\item  Real-time state estimation and major system frameworks supporting the real-time management system are analyzed.
\item System performance improvement methods and emerging topics are reviewed.

\end{highlights}

% Keywords

\maketitle

\section{Introduction}

As industrial society continues to evolve, the transportation sector has remained a major contributor to greenhouse gas (GHG) emissions. While traditional internal combustion engine vehicles have facilitated convenient mobility, they have also significantly polluted the environment. In recent years, as the transportation sector strives for cleaner and more sustainable solutions, electric vehicles (EVs) have become a key factor in reducing GHG emissions \cite{mobarak2020solar}. With their low maintenance requirements and superior performance, the number of EVs has increased rapidly, marking significant progress in the electrification of transportation \cite{fachrizal2021combined}. According to the International Energy Agency's Global EV Outlook 2024, the global stock of EVs, excluding two- and three-wheelers, is projected to grow from under 45 million in 2023 to 250 million by 2030, and further to 525 million by 2035. By that time, more than a quarter of all vehicles on the road will be electric. The widespread adoption of EVs offers numerous benefits, such as reducing GHG emissions through decreased fossil fuel use and enhancing transportation capacity to lower costs. However, integrating EVs on a large scale into distribution networks poses significant challenges for grid operation. It is anticipated that by 2040, EVs will account for approximately 28\% of the market share, resulting in an 11-20\% increase in global electricity consumption \cite{kapustin2020long}.
Additionally, the increased demand from EV charging during peak hours could place greater pressure on the safe and stable operation of existing distribution system. This could also result in higher cost, as consumers are often required to pay substantial charges based on time-of-use tariffs \cite{engel2022hierarchical}. To address these challenges, it is crucial to actively integrate EVs into energy infrastructure and management systems. With their dual characteristics serving as controllable loads or energy storage devices \cite{li2021coordinating}, EVs can help smooth load fluctuations in the power grid, balance the intermittency of distributed generation (DG), and provide ancillary services to maintain grid stability. Therefore, through appropriate real-time management strategies, the negative impacts of EV charging can be minimized, or even transformed into positive outcomes. In this regard, this paper focuses on real-time EV charging management. 

Recent review articles have extensively discussed the impacts of EV charging on the grid. They generally focus on specific research directions, including  V2G technology \cite{ibrahim2024analysis}, EV-grid interaction \cite{motlagh2025review}, charging topologies \cite{rahman2022comprehensive}, charging technologies \cite{rana2025comprehensive}, power quality issues \cite{inci2024power,rahman2022comprehensive,khalid2019comprehensive} and mitigation approaches \cite{srivastava2023electric,wang2021grid,ahmed2026holistic}, EV charging and battery swapping infrastructure \cite{suresh2026comprehensive, zhan2022review, wu2021survey}, cybersecurity \cite{ronanki2023electric}, and EV technologies \cite{wang2023review, das2020electric,khalid2021comprehensive}. Nevertheless, existing review articles have not sufficiently addressed real-time EV charging management as a systematic research topic. Motivated by this gap, this paper provides a comprehensive review of real-time EV charging management and its related issues. This paper first illustrates the negative impacts of EV charging on the real-time operation of distribution systems. It then focuses on mitigating these impacts from two perspectives:
\begin{itemize}
    \item[1)] Providing a comprehensive summary of real-time grid management strategies to regulate EV impacts, such as smart charging, charging environment management, energy coordination, battery management, and ancillary services. 
    \item[2)] Conducting an in-depth analysis of the real-time management system for EV integration by focusing on real-time state estimation of the distribution networks, system frameworks for the management system, emerging technology integration, and future research directions.
\end{itemize}

\begin{figure*}[h]  
    \centering      \includegraphics[width=1\textwidth]{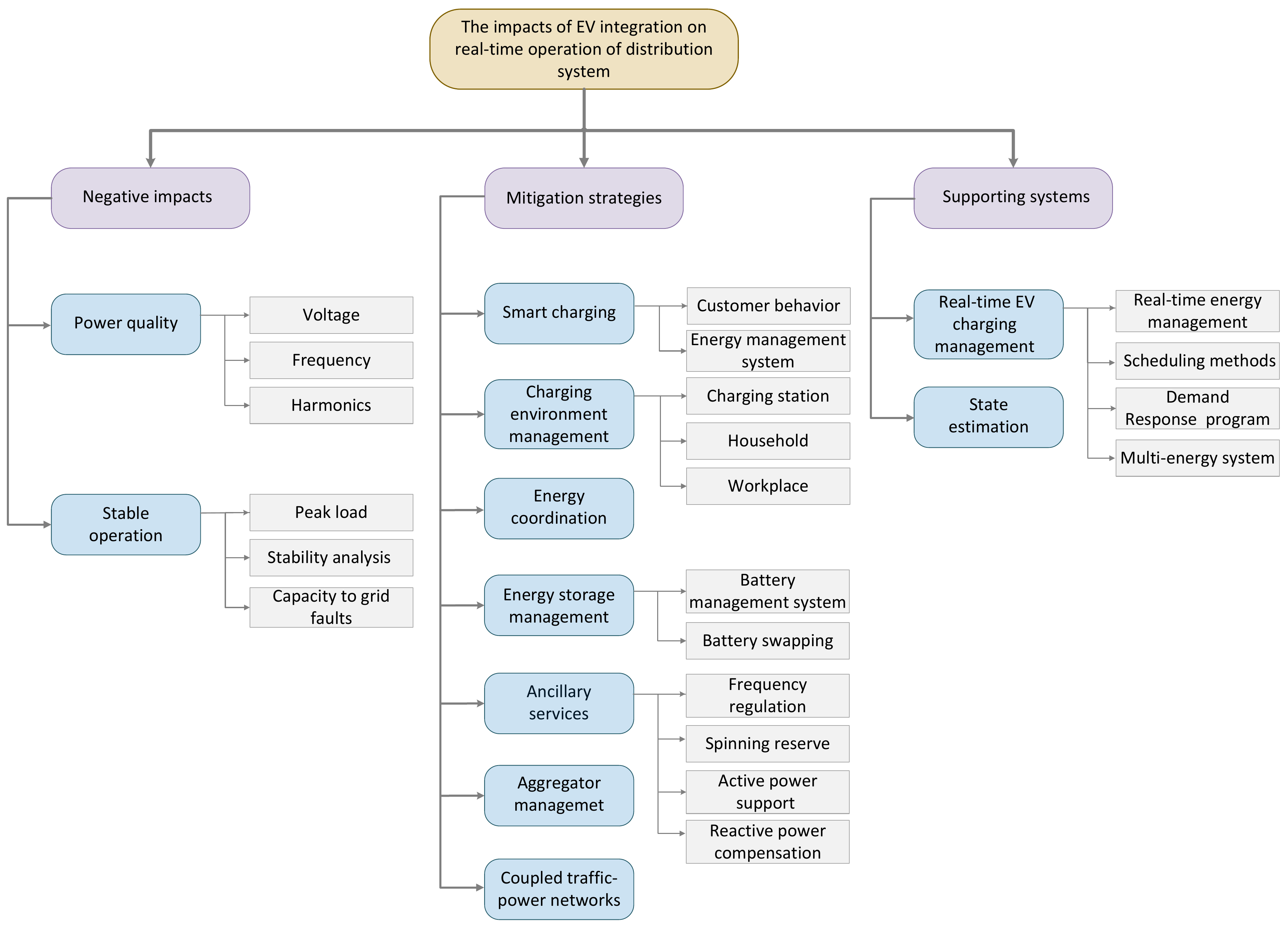}  
    \caption{Summary of research on EV charging's impacts reviewed in this study.}  
    \label{fig: 1}  
\end{figure*}

The overall structure of this paper is depicted in Figure \ref{fig: 1}. The rest of this paper is organized as follows. Section 2 discusses the negative impacts of EV integration on the real-time operation of distribution systems. Sections 3 and 4 examine the grid management strategies and the real-time EV management system to mitigate these impacts, respectively. Finally, Section 5 offers conclusions and insights for future research. 

\section{EV integration impact on real-time operation of distribution systems}
\subsection{Power quality} 
\subsubsection{Voltage} 

Based on the characteristics of the power system, voltage is closely related to reactive power, which is more difficult to generate and inject into grids than active power \cite{sayed2022electric}. EV charging is a typical non-linear load based on rectifier circuits and power converters. Thus, the high demand for reactive power from non-linear  EV charging can significantly affect the real-time voltage profile of grid operation, which reduces the power factor and exacerbates voltage distortion and fluctuations. In addition to the increase in reactive power demand, large-scale EV charging and the deployment of high-power chargers draw substantial current from the grid, leading to increased active power transfer. Transmitting large amounts of reactive and active power from generators to loads results in significant power losses in distribution lines. In power grids, the voltage magnitude difference and phase angle difference between two nodes are directly related to transmission losses \cite{hasan2023distribution}. Charging location, transmission distance, charging power demand, and EV penetration level are among the main factors influencing voltage variation. Thus, the direct impact of large-scale EV integration is a voltage drop at the coupling point, which may lead to voltage deviation exceeding the regulatory requirements. Review \cite{ahmed2026holistic} summarizes the literature that tested and quantified the impact of EV charging on voltage profiles in distribution networks. The results show that the difference in the voltage deviation index between 50\% EV penetration and 20\% EV penetration is 8.3\%. Under normal operating conditions, 100\% EV penetration can alter the voltage of medium-voltage networks by approximately 0.02–0.05 p.u., while the voltage variation in low-voltage networks can reach approximately 0.10–0.15 p.u. \cite{rahman2022comprehensive} investigates the factors affecting voltage distribution, such as locations of power sources and EV penetration levels, and compares conditions of multiple parallel load lines with unequal loads. Furthermore, continuous violations may lead to grid operation instability and could even result in blackouts. Therefore, when voltage violations or frequency violations exceed specified limits, it’s necessary to adopt corrective measures to restore violations to normal levels in order to avoid damage to power equipment and negative effects on grid operation safety. 

Regarding voltage regulation, reactive power dispatch and load demand management are effective methods to control voltage drops \cite{wu2012pev, leemput2015reactive, sousa2015multi}. Some studies \cite{zipeng2017robust,li2012impacts} compare the effects of random charging and smart charging on the voltage variation of the distribution system, finding that smart charging can increase EV penetration rates. Moreover, another impact on voltage is the issue of three-phase voltage unbalance, which is primarily caused by single-phase charging. Uneven distribution of charging loads across three phases may lead to a severe condition of three-phase voltage unbalance \cite{yong2015review}.  \cite{richardson2010impact} performs a detailed study on this issue by connecting all EVs to a single phase and confirms the seriousness of the problem. To address this problem, a proper load management strategy is usually adopted to mitigate three-phase voltage unbalance by evenly dispatching EV charging loads across three phases.  

%\begin{figure}[h]  
    %\centering      %\includegraphics[width=0.46\text%width]{051601.pdf}  
    %\caption{The phasor diagram %for regulating voltage deviation %to the normal level.}  
    \label{fig:2}  
%\end{figure}

\subsubsection{Harmonics}

Power electronic devices are extensively used in EV charging and discharging systems, which may introduce power quality issues, particularly harmonic distortion, and consequently affect the real-time operation of distribution systems \cite{yong2015review}. The architecture of the EV chargers is responsible for the insertion of particular order of harmonic in the grid, which are typically odd-order harmonics \cite{khalid2019comprehensive}. Total harmonic distortion (THD) can be used to quantify the effective value of the harmonic contents of a distorted waveform \cite{srivastava2023electric}. \cite{senol2024harmonics} measured the harmonic insertions of eight different EV models under various smart charging rates. The results show that the THD values are lowest when the vehicles are charged at the maximum rate. Regarding the relationship between individual current harmonic orders and charging current, the 7th, 3rd, and 5th harmonic components exhibit larger amplitudes  among all tested vehicles, regardless of the charging current level. By further analyzing the harmonic phase angles, a certain degree of harmonic cancellation among different vehicles is observed due to 180-degree difference between their phase angles. Regarding studies on the effect of harmonics, establishing models of EV chargers is a critical point, where \cite{gatta2016pq} develops a sensitivity analysis about the composition of harmonic disturbances due to AC/DC converters installed in EVs, and \cite{rahman2022comprehensive} compares simulation results for different power converter topologies in steady-state operation. The basic configuration of an EV charger uses a back-to-back converter structure. On the input side, a diode bridge rectifier determines the current, which is highly peaked and dominated by low-order harmonics. This leads to voltage deviations, reduced power quality, and de-rating of system components \cite{yong2015review,rahman2022comprehensive}. The peak current is superimposed on the sinusoidal current drawn by the EV charger and other loads in the distribution system. It produces a non-sinusoidal voltage drop across the grid impedance. Thus, both the coupling point and the distribution grid contain additional harmonics. The effect of the above condition depends on the parameters of the distribution line. If the grid impedance is small, the voltage drop at the coupling point due to non-sinusoidal current is small. Although EV charging may bring harmonic pollution to the power grid, employing filtering and advanced power electronic devices can alleviate this problem. For instance, \cite{rai2019bridgeless} proposes a single-ended primary inductor converter for power factor correction operation, and \cite{rahman2022comprehensive} adopts a boost power factor correction circuit along with the diode bridge rectifier to solve the above condition, through improving and regulating the rectified voltage to generate a minimal ripple DC voltage. Future studies should also incorporate the influence of background harmonics, which would help clarify the interaction between EV-induced harmonics and existing harmonic components in distribution networks \cite{senol2024harmonics}.

\subsection{Stable operation of distribution systems}
Large-scale uncoordinated charging of EVs may negatively affect the real-time operation of the distribution system, causing overloading, voltage drops, power outages, and posing a threat to the stable and safe operation of the distribution network \cite{hasanien2015adaptive}. This stable problem can be written as: 

For any initial state $x(t_{0})$ with $\left\|x\left(t_{0}\right)\right\|<k_{1}$  and any input  $u(t_{\mathrm{s}})$  with  $\sup _{t_{\mathrm{s}} \geq t_{0}}\|u(t_{\mathrm{s}})\|< k_{2}$,  $x(t_{\mathrm{s}})$ exists and the output $y{(t_s)}$ satisfies
\begin{eqnarray}
    \label{e2}
    \begin{gathered}
\|y(t_{\mathrm{s}})\| \leq \beta\left(\left\|x\left(t_{0}\right)\right\|, t_{\mathrm{s}}-t_{0}\right)+\alpha\left(\sup _{t_{0} \leq \tau \leq t_{\mathrm{s}}}\|u(\tau)\|\right),
    \end{gathered}
\end{eqnarray}
where all  $t_{\mathrm{s}} \geq t_{0} \geq 0$.

Research on the impact of EV charging on the stability of the distribution system typically focuses on three aspects: rotor angle stability, frequency stability, and voltage stability. References \cite{dharmakeerthi2014impact,shi2012dynamic,zhou2016assessment} indicate that EV charging reduces the level of power system stability, while references \cite{wu2011transient,liu2021optimal} suggest that the Vehicle-to-Grid (V2G) model can improve the system stability level.

\subsubsection{Peak loads}

Both electricity consumption and EV charging loads are closely related to human activities \cite{wang2022cyber}. Without controlled charging, EV charging demand may overlap with existing peak loads in the grid \cite{fachrizal2020smart}. This overlap would further intensify grid loads during peak time, thereby stressing grid operation and posing risks to the security and stability of the distribution system. Peak load increase generated by EV charging has become a critical factor for grid operation and risk assessment \cite{stiasny2021sensitivity}. Some studies have focused on this issue. For instance, \cite{weiller2011plug}  suggests that, during commuting hours, new load peaks could exceed natural peaks if EV charging loads are not sufficiently shifted to off-peak periods. Another study \cite{shafiq2020reliability} indicates that uncontrolled EV charging, especially during peak time, could lead to up to 6.89\% load loss. Since traditional distribution systems are designed to handle peak loads, reducing peak demand can also significantly lower overall construction costs. Key factors affecting peak loads in the distribution system include EV charging time, charging location, charging power, and penetration rates of EV charging. Thus, the load management strategy can help balance power loads and reduce the difference between peak and valley loads \cite{fachrizal2020smart}. Common load management methods include off-peak and valley-filling charging, which shifts EV charging from peak to lower-demand times. This charging approach avoids charging during peak periods and fills low consumption periods, reducing system loss and improving load factor \cite{sortomme2010coordinated}.

Regarding the impact of EV-grid-connected charging on peak loads in practical applications, many countries, including the United States and Germany, have analyzed the effects of EV charging on load distribution based on their specific circumstances and have proposed corresponding solutions. These solutions include delayed charging methods \cite{weiller2011plug}, using EVs as stable power storage devices in the grid \cite{hartmann2011impact}, shifting EV charging to nighttime hours \cite{mullan2011modelling}, transferring EV loads from peak to off-peak periods by implementing demand response (DR) strategies \cite{park2013impact}
and utilizing V2G reverse power flow to reduce peak load in the grid \cite{hu2019prediction}. Specifically, \cite{shao2012grid} establishes a real-time energy management optimization model for an EV parking lot based on a peak load limitation oriented DR program  to maximize the load factor. The simulation results in the distribution circuit in Blacksburg show that the proposed DR strategy can maintain the original peak demand with different EV penetration levels. In \cite{binetti2015scalable}, a scalable real-time greedy algorithm is used to coordinated charging strategies, which reduces the peak value to 10709 MW, compared with the base profile peak of 16327 MW. \cite{hu2019prediction} simulates V2G mode to reduce peak load on a real low voltage network in England, where the power curve is levelled off at 20\% penetration and the maximum penetration level is 50\%. \cite{mosaddegh2017optimal} developed a distribution optimal power flow model incorporating a neural network model of controllable loads to mitigate peak loads. Based on a microgrid energy management system framework, \cite{solanki2015including}
proposes optimal dispatch strategies for dispatchable generators, energy storage systems, and controllable peak loads to achieve peak load dispatch effectively. 

\subsubsection{The coping capacity of EV charging areas to grid faults} 

Faults in grids may cause fast variations of voltage and bursts of harmonics, as well as a serious influence on EV charging. For instance, a fault causing undervoltage at or below 0.3 p.u. can cease charging for 2–10 seconds \cite{mckillop2023impact}. \cite{onar2010grid} compares conditions of a three-phase ground fault and after fault clearance, obtaining the conclusion that systems connected to EVs are more sensitive to disturbances and less stable in magnitude deviations and adjustment time. \cite{mckillop2023impact} investigates the impact of EVs during network faults through testing EVs' responses to a double-line-to-ground transmission fault, obtaining the result of successive over-frequency and under-frequency of 50.78 and 49.22 Hz, respectively. Therefore, regarding fault conditions in grids, the ability of EV charging areas to ride through grid faults and mitigate grid faults is worth attention. EV charging areas with low-voltage ride-through (LVRT) function can effectively handle faults and prevent system instability \cite{onar2010grid}, and V2G mode can also be used for grid support during faults according to LVRT requirements \cite{katic2019impact}. For instance, \cite{katic2019impact} tests six characteristic types of faults to demonstrate the positive impacts of LVRT. Table \ref{tab1} summarizes the research on enhancing EV charging areas' coping capacity to grid faults.

\begin{table*}[t]
\caption{Summary of the research on enhancing EV charging areas’ coping capacity to grid faults.} 
\begin{tabular}{>{\centering\arraybackslash}p{0.005\linewidth}>{\centering\arraybackslash}p{0.005\linewidth}>{\raggedright\arraybackslash}p{0.14\linewidth}>{\raggedright\arraybackslash}p{0.23\linewidth}>{\raggedright\arraybackslash}p{0.15\linewidth}>{\raggedright\arraybackslash}p{0.18\linewidth}}
\hline
\multicolumn{1}{c}{Ref. no.} &
  \multicolumn{1}{c}{Year} &
  \multicolumn{1}{c}{Technical foundation} &
  \multicolumn{1}{c}{Implementation approach} &
  Aims and resolved issues  &Performance metrics\\ \hline
\cite{mishra2022intelligent}&
  2022 &
  Least mean  square algorithm &
  A reconfigurable multi-objective charging control architecture &
  Overcome various grid abnormalities during EV charging operation  & The LVRT operation is shown during 0.7-0.8s\\
\cite{baghaee2019anti} &
  2020 &
  Support vector machines &
  Anti-islanding protection scheme for low voltage-sourced   converter-based microgrids &
  Islanding and grid-fault detection.  &Islanding detection is achieved
within 45-60 msec\\
\cite{shin2012development} &
  2012 &
  Wireless sensor network &
  Smart grid monitoring system connected with EV charging system using anti-islanding method &
  EV charging process continues without any serious fault  &The operation of micro-grid system is performed well\\
\cite{onar2010grid} &
  2018 &
  Dynamic combination of EV chargers and single-phase induction motors &
  Implement an LVRT scheme to inject reactive power into the grid and regulate EV charging rate &
  Handle faults and prevent dynamic voltage instability.  &When the voltage becomes 
stable, EV loads can be charged at full 
rate after about 0.6 seconds\\
\cite{almeida2015electric} &
  2015 &
  Combination of inertial  emulation and droop control &
  Primary frequency control  technique with EVs &
  Safe integration of intermittent renewable energy sources  &It is verified that EV participation in frequency control in an
isolated test system reduced the frequency oscillation band of the
system\\
\cite{pandit2024frequency} &
  2024 &
  MM-SFR model &
  Propose a  non-linear optimization framework incorporating constraints of EV aggregator, frequency security, converter voltage, security, and LVRT constraints &
  Develop a framework for quantifying EVs' contribution to providing frequency support  &Computation time of  around 3 s/step\\
\cite{mckillop2023impact} &
  2023 &
  Fault condition test &
  Test EVs physically under various network fault conditions using a grid simulator supply interface &
  Characterize fault ride-through (FRT) performances  &The aggregation of EV fault-responses yields a resultant successive over-frequency and 
under-frequency of 50.78 and 49.22 Hz respectively\\
\cite{katic2019impact} &
  2019 &
  V2G mode according to the LVRT requirements &
  Test the possibility of voltage improvement during voltage dips in V2G mode &
  Provide grid support during faults through V2G mode  &65.7\% higher voltages can be 
achieved during the dip with LVRT supported by EV\\
\cite{yadav2023low} &
  2023 &
  LVRT &
  Applications of LVRT &
  Boost the resilience of the power  network against extreme events  &With the inclusion of the fault impedance, the grid
voltage does not collapse to zero\\
\cite{falahi2013potential} &
  2013 &
  V2G services &
  Pair up a photovoltaic source with an EV charger through a single-phase bidirectional charger topology &
  Study the  potential of EVs to help PV sources during LVRT  &The
charger can keep the
voltage at the nominal value\\
\cite{dietmannsberger2017simultaneous} &
  2017 &
  A three-phase inverter  model &
  Inverters with both   LVRT capability and  anti-islanding protection simultaneously &
  Solve the conflict between LVRT capability and anti-islanding detection requirements  &The inverter can
behave correctly in all necessary cases, even unbalanced conditions\\ \hline
\end{tabular}
\label{tab1}  
\end{table*}

\section{ Real-time management strategies }
The impact analysis in Section 2 shows that large-scale EV charging may introduce several operational challenges to distribution systems, including power quality deviation and stable operation risks. Many research works focus on the mitigating methods to address these problems. This paper aims to reverse these negative impacts to positive effects by applying grid management methods. This section summarizes the mitigating methods from the perspective of real-time grid management, including smart charging, charging environment management, energy coordination, battery management, and ancillary services. 

\subsection{Smart charging}

\begin{figure*}[htbp]
  \centering
  \begin{subfigure}[b]{0.4\textwidth}  \includegraphics[width=\textwidth]{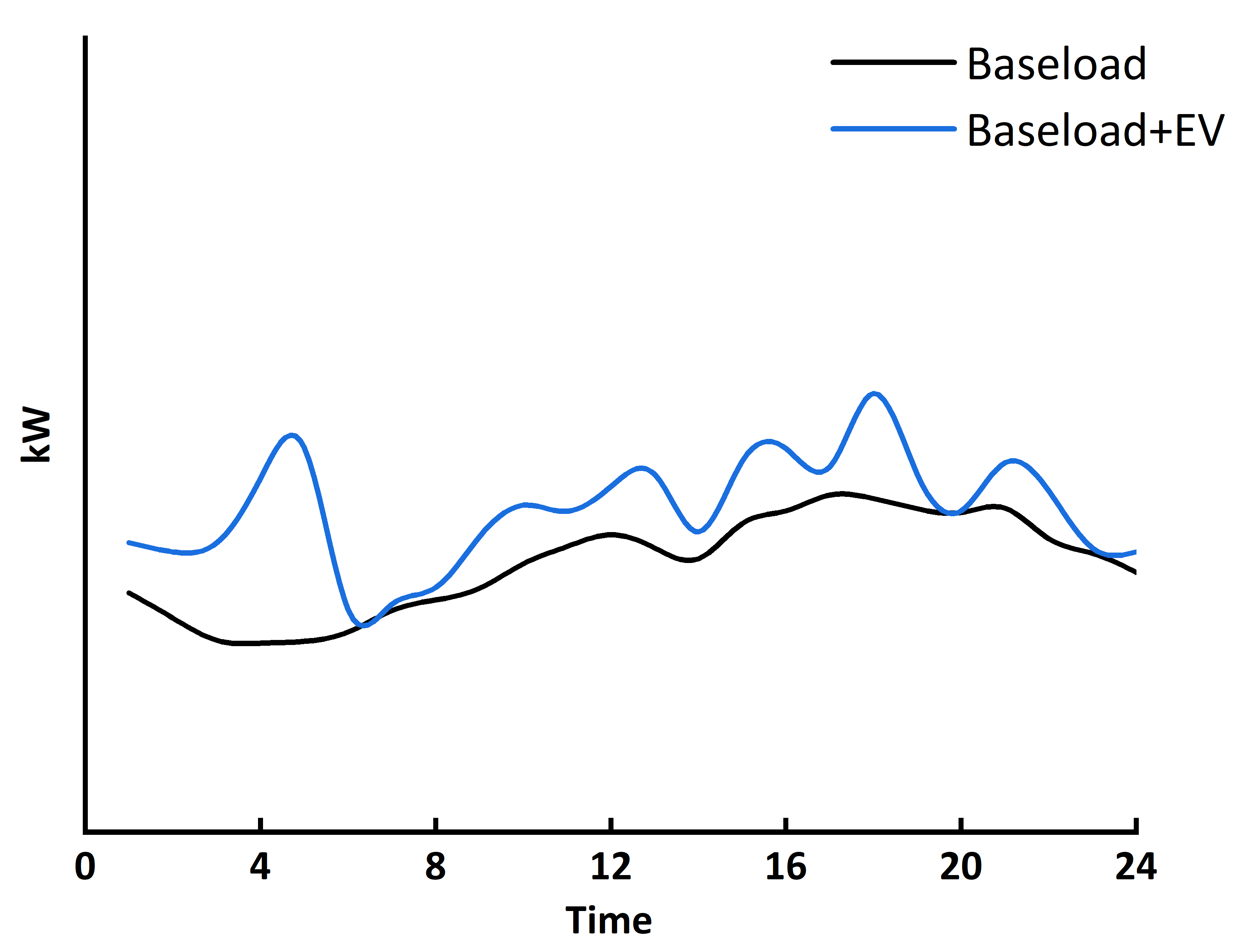}
    \caption{ }
    \label{fig:sub1}
  \end{subfigure}
  \hfill
  \begin{subfigure}[b]{0.4\textwidth}
  \includegraphics[width=\textwidth]{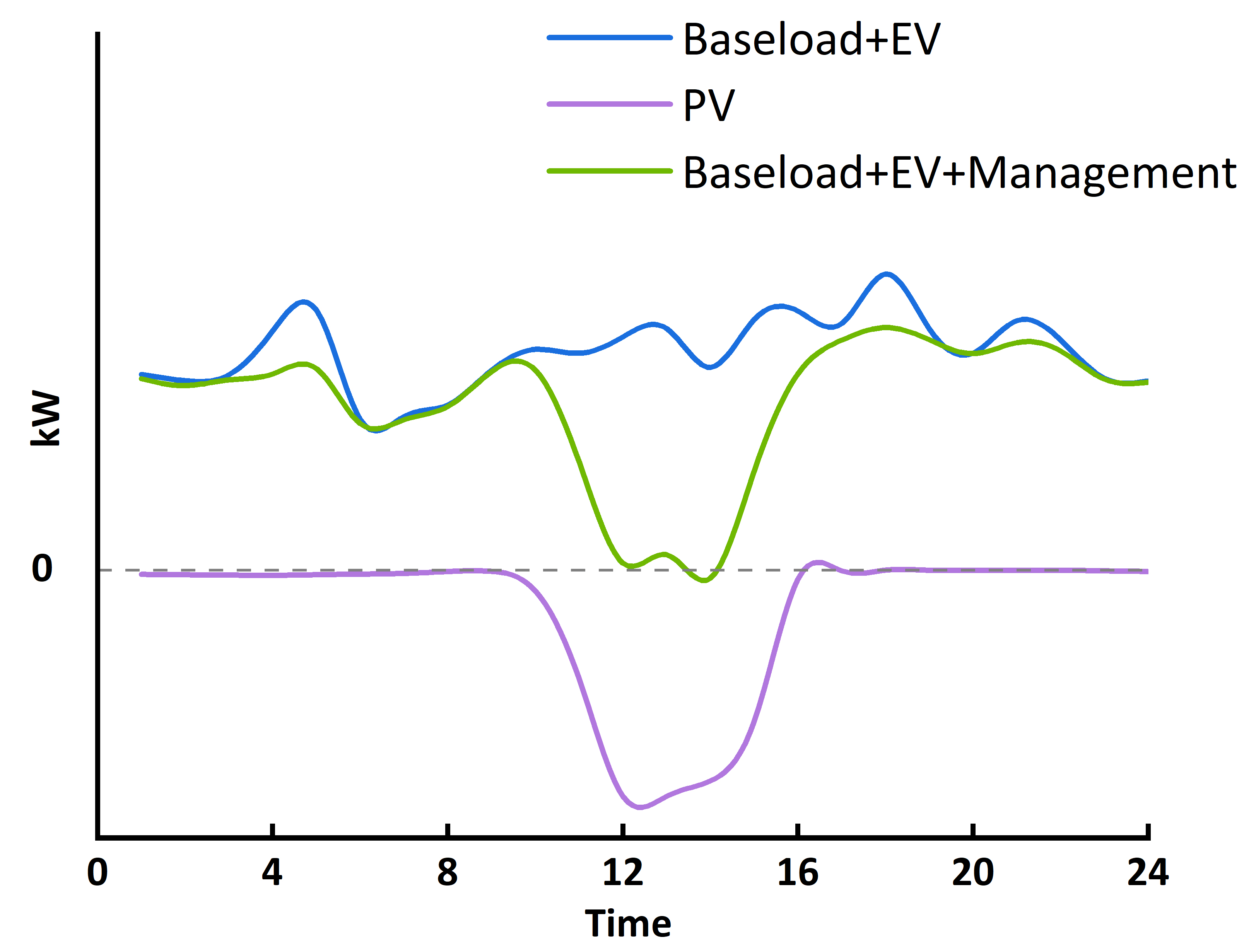}
    \caption{}
    \label{fig:sub2}
  \end{subfigure}
  \caption{(a) The effects of EV charging on the electricity load profile on a day in residential distribution grids. (b) The positive impacts of managing EV charging on net loads through PV coordination and smart charging methods. Figures are developed with data from \cite{berg2022data}.}
  \label{fig:main}
\end{figure*}

In response to the growing number of EVs and their impacts on infrastructure, it is necessary to implement intelligent and coordinated charging management methods \cite{engel2022hierarchical}. Smart charging refers to the intelligent scheduling of EV charging by leveraging data and communication technologies to reduce the adverse impacts of uncontrolled charging, considering grid conditions, electricity prices, and user travel needs \cite{dahiwale2024comprehensive}. Many studies have focused on optimizing charging management through smart charging approaches, typically integrated with an energy management system (EMS), which is designed to determine optimal charging schemes and regulation strategies that lead to positive effects on the distribution network.  Figure \ref{fig:main} presents the negative impacts from EV charging on net loads, and the positive impacts from smart charging. Through establishing models for controllable loads \cite{solanki2015including,hafez2016integrating,mosaddegh2017optimal}, these models are integrated into the distribution system operation framework or EMS \cite{solanki2017sustainable,wi2013electric,van2016energy,solanki2015including} to determine the optimal charging strategies and dispatch decisions \cite{wi2013electric,van2016energy,engel2022hierarchical,solanki2015including,hafez2016integrating,kiviluoma2011methodology,zheng2018online}. The advantage of this approach is optimizing EV charging management together with other components of power system. Determining the optimal charging strategies and optimal dispatch decisions is typically formulated as an optimization problem considering multiple factors, with the charging management objective encoded as a cost function \cite{engel2022hierarchical}. Specifically, in \cite{solanki2015including}, an EMS framework is proposed to determine optimal scheduling decisions considering dispatchable generators, energy storage systems, and peak demand for controllable loads. \cite{zheng2018online} transforms the optimal charging problem to an optimal power flow problem to minimize the total system energy cost, then utilizes a modified convex relaxation technique to obtain the globally optimal solution. The key issue in establishing EV load models is reducing the uncertainties caused by EV charging behaviors. Due to the complexity of controllable loads and the limited data, it is difficult to model these loads using basic physical laws \cite{mosaddegh2017optimal}. Common methods include probabilistic approaches \cite{papadopoulos2010predicting,ramadhani2020review}, stochastic optimization \cite{van2016energy,liao2015dispatch}, and evolutionary algorithms, such as neural networks \cite{hafez2016integrating,mosaddegh2017optimal}, particle swarm optimization \cite{regulski2014estimation}, or genetic algorithms \cite{ju1996nonlinear}. Moreover, based on these technical foundations, many studies have focused on developing efficient and accurate real-time management models to address the impact of EV integration on the real-time operation of distribution networks. A detailed introduction to this topic is provided in Section 4. 

\subsection{Charging environment management}

In addition to charging stations, common EV charging environments include workplaces and households, where rooftop photovoltaic (PV) systems are typically integrated with energy storage systems (ESS). Regarding workplace conditions, the net-zero energy building (NZEB) is a popular topic, which aims to ensure that on-site electricity generation can fully meet the building’s total electricity demand. While the concept of building energy management system (BEMS) is traditionally applied to control heating, ventilation and air conditioning  (HVAC) systems and determine operating schedules in order to reduce energy consumption, NZEB requires a more integrated approach where different types of energy sources are interconnected within the building and coupled with the power grid. Typically, this issue begins with modeling energy-efficient building by energy optimization analysis, solar energy and EV batteries are further integrated into the building energy system, followed by optimization analysis for the best design alternative. Three main aspects are mainly considered: total site energy consumption, capital cost, and comfort level \cite{alirezaei2016getting}. Identifying the optimal design can be challenging due to conflicting objectives, thus multi-objective optimization methods are often required. For example, \cite{alirezaei2016getting} investigates the issue based on a system of PV panels, EVs, the main battery, and the power grid. Among nearly 1,990 setpoints, 6 points are selected as optimal alternatives. EVs are modeled and incorporated into the energy system as mobile batteries during non-working hours. Comparative simulations demonstrate that this system could reduce grid electricity demand by up to 68\%, and lower electricity bills by 62\%. \cite{engel2022hierarchical} proposes a hierarchical economic model predictive control scheme for EV charging management, considering the objectives of monetary costs, building temperature comfort, EV charge satisfaction, and battery degradation. \cite{van2016energy} designs a workplace energy management system with photovoltaic generation prediction and power flow optimization between PV systems, grids, and battery electric vehicles. \cite{lakshminarayanan2018real} proposes a real-time optimal EM controller for V2G integration to provide an optimal schedule for the operation of the workplace microgrid system.

The issue of EV charging in households is closely related to the home energy management system (HEMS), which is necessary for residential electricity consumers to participate in DR programs actively \cite{ghazvini2017demand}. This issue is typically formulated as an optimization problem, often modeled using a Markov decision process. The objective is to minimize the occupant’s utility function while considering constraints at multiple levels, such as occupant, residential home and distribution grid levels \cite{liu2021pv}. However, due to the difficulty in accurately quantifying occupant behavior, the effectiveness of proposed strategies is highly dependent on assumed scenarios. For example, \cite{seal2023centralized} applies a centralized model predictive control (MPC) strategy with zone-based control to manage a heating system comprising a heat pump with multi-split fan units and electric baseboards while integrating PV generation and EV energy storage. The responsiveness of the MPC is evaluated in a vehicle-to-home (V2H) case study where EV arrival time is only notified a few minutes before arrival. \cite{garifi2019stochastic} proposes a chance-constrained MPC algorithm to manage controllable resources, including PV panels, home batteries, EVs, and HVAC systems, with the goal to ensure indoor thermal comfort despite uncertainties in temperature and solar irradiance forecasts. \cite{alirezaei2016getting} compares three scenarios differing in EV energy operations and PV placements, and finds the grid electricity consumption can be reduced up to 45\% and 77\%. \cite{irfan2024novel} develops a base case to analyze grid power-sharing based on a grid-assisted bidirectional PV–EV system using the system advisor model and conducts tests in Sydney households. \cite{liu2021pv} develops a stochastic adaptive dynamic programming model to optimize HVAC setpoints, clothing behavior, and EV energy scheduling, accounting for uncertainties in outdoor temperature, PV generation, and EV’s state of charge (SOC).

Some studies investigate the optimal energy flow that motivates these scenarios, focusing on balancing thermal comfort, electricity cost minimization, and the integration of distributed energy resources. For instance, \cite{seal2023centralized} develops an MPC-based HEMS to manage zone-based thermal comfort along with optimizing the energy flow among the components of the home energy network. Specifically, the MPC optimizes heating system inputs to minimize energy cost, including the part load ratio and the percentage of the rated baseboard heating input. A multistep MPC feedback strategy is employed with a simulation time step of 3.75 min, a prediction horizon of 8 h, and a control horizon of 15 min. The MPC also achieves approximately 8\% reduction in the energy cost compared with the base-case scenario. \cite{irfan2024novel} proposes a data-driven HEMS model based on the proximal policy optimization algorithm to optimize policy formulation in sequential decision-making tasks. It prioritizes the lowest-cost energy sources and leverages V2H and V2G functionalities to minimize monthly electricity costs. In \cite{van2016energy}, an EMS that combines an autoregressive integrated moving average model for PV forecasting and a mixed-integer linear programming framework for power allocation is employed to optimize power flows among PV, EV, and the grid. This approach reduces EV charging costs by 118.44\% with one charging point and 427.45\% with two charging points compared with uncontrolled charging. \cite{liu2021pv} introduces HEMS based on adaptive dynamic programming. A model predictive control framework is further integrated to continually update optimal appliance scheduling decisions under time-of-use tariff, achieving a 68.5\% reduction in energy costs compared with conventional scheduling strategies. 

%%%%%%%%%%%%%%%%%%%%%%%%%%%%%%%%%%%%%%%%%%%%%%%%%%%%%

\subsection{Energy coordination and battery management}

Renewable energy generation is flexible, environmentally friendly, and cost-effective, significantly reducing greenhouse gas emissions and environmental pollution. It is produced naturally and is subject to natural laws, which means it is inherently random and discontinuous. For example, wind power generation depends on variations in wind speed and direction, while photovoltaic generation is influenced by solar irradiance and shadow patterns, which are affected by geographical location, micro-climates, and seasons. Consequently, renewable energy generation is intermittent and fluctuating, posing challenges for power systems to maintain a real-time balance between supply and demand, thus requiring effective management. Additionally, since both electricity consumption and EV charging loads are closely linked to human activities, peak power consumption and EV charging demands are likely to occur simultaneously \cite{fachrizal2020smart}. Distributed generation (DG) is often unstable and difficult to predict accurately. Coordinated EV charging can help balance power load demands at various times, thereby mitigating the intermittent and unstable effects of DG \cite{ghazvini2017demand,wi2013electric,van2016energy,solanki2015including,shepero2018modeling}. This problem can be written as \cite{ramadhani2020review}:
\begin{align}
\min & \sum_{t=t_{\mathrm{arr}}}^{t_{\mathrm{dep}}}\left(p_{t}+l_{t}-s_{t}-\mu_{\mathrm{tpark}}\right)^{2}, \\
\text { s.t. } & \eta_{p} \sum_{t=t_{\mathrm{arr}}}^{t_{\mathrm{dep}}} p_{t} \cdot \Delta t=\mathrm{SoC}_{\mathrm{tar}}-\mathrm{SoC}_{\mathrm{arr}}, \\
& 0 \leq p_{t} \leq p_{\max },
\end{align}
where  $t_{\text {arr }}$  and  $t_{\text {dep }}$  are the arrival and departure times of EV respectively,  $p_{t}$  is the charging power rate at time  $t$, $l_{t}$  is the household load at time $t$, $s_{t}$  is solar power generation rate at time $t$, $\mu_{\mathrm{tpark}}$  is the mean net-load. In the constraint, $\eta_{p}$ is the charging efficiency, $\Delta t$ is the time step, $\mathrm{SoC}_{\mathrm{tar}}$ is the targeted state of battery, $\mathrm{SoC}_{\mathrm{arr}}$ is the state of battery at arrival time and $p_{\max }$ is the maximum charging power rate.

An electric vehicle represents a movable battery storage load \cite{sayed2022electric}, which can be used as a flexible and mobile storage unit. Given the high cost of energy storage systems, enhancing the utilization of DG through the control of flexible loads represents one of the effective solutions. To effectively guide EV users toward expected charging behaviors, \cite{chen2020blockchain} proposes a blockchain-based EV incentive system that incorporates a prioritization ranking algorithm for EV drivers to maximize the utilization of renewable energy, which can be deployed at a large scale. Focusing on the coupled stochastic effects of wind-solar generation and user charging preferences, \cite{chung2020intelligent} introduces two stochastic game models to investigate the complex interactions between the power grid and charging stations and to model the dynamic variations in users’ preferences for charging parameters. An online algorithm is then developed to solve the Nash equilibria of the stochastic games. The results show that the proposed method can reduce electricity costs by approximately 20\%.

Two main types of electric vehicle supply equipment (EVSE) services are currently available in the market. In addition to conventional plug-in charging services, battery swapping stations (BSSs) can serve as an alternative solution to overcome the limitations of charging speed in plug-in charging stations. Owing to their ability to provide rapid and sustainable battery replacement services, BSSs are particularly suitable for two types of commercial vehicles \cite{chen2021electric}: heavy-duty commercial vehicles and taxis or commercial passenger vehicles.  The comparison among plug-in charging station, mobile charging station and BSS is presented in Table \ref{tab2}. In \cite{zhang2018monte}, Monte Carlo simulations are used to compare the service capacities and profitability of charging and battery swapping services for electric taxi and electric bus fleets. The results show that, under the same service capacity, BSSs have greater economic potential than conventional charging stations, especially for vehicle operators and EVSE service providers. To further quantify the interactive dynamics among EVs, EVSE services, and user behaviors, as well as the overall benefits generated for users and service providers, \cite{zhang2021stochastic} proposes a discrete stochastic model for interactions between CS/BSS and taxi/bus fleets. The revenues of the stochastic service system are then derived and verified by simulations.

The large-scale energy storage capacity of BSSs can provide substantial support for grid operation. However, the grid-connected charging of a large number of batteries may also aggravate several grid-side issues, such as voltage instability, harmonic distortion, and peak load superposition. Therefore, appropriate charging locations, priority charging strategies for depleted batteries, and interaction mechanisms between BSSs and the power grid are critical for the efficient operation of BSSs. For example, charging batteries within pre-scheduled time periods can reduce EV charging uncertainties \cite{amiri2018multi}. Charging batteries during off-peak periods can also alleviate peak-load pressure on the grid. In addition, coordinated battery charging in BSSs can help mitigate power quality issues, such as harmonics \cite{zhan2022review}.
BSSs can also participate in V2G frequency regulation by reserving a large amount of regulation capacity through stored batteries. In contrast, plug-in electric vehicles (PEVs) usually need to be centrally managed through aggregators to accumulate sufficient capacity for frequency regulation participation. BSSs can respond to area control error signals in real time, whereas hierarchical communication networks between PEV aggregators and individual PEVs may introduce communication delays. However, BSS-based V2G also requires higher infrastructure investment, such as hybrid AC-DC/DC-AC inverters to support bidirectional energy flow, and needs to account for increased battery aging costs and uncertainties in battery charging costs. To address these issues, \cite{wang2020economic} and \cite{wang2020vehicle} propose a BSS-based economic risk assessment model for fast frequency regulation services based on reinforcement learning and a dynamic scheduling strategy based on deep Q-learning, respectively. \cite{wang2024optimal} proposes a day-ahead bidding and real-time scheduling strategy for BSS participation in frequency regulation. However, the standardization of battery packs and the sharing of battery technologies remain key challenges for the large-scale deployment of BSSs. Future research should pay attention to battery standardization, the optimization of charging and swapping processes, the extension of battery health, and operational models that consider battery heterogeneity \cite {wu2021survey}. 

\begin{table}
    \centering
\caption{The comparison of EVSEs.}
    \label{tab2}
    \begin{tabular}{|c|>{\centering\arraybackslash}p{0.12\linewidth}|>{\centering\arraybackslash}p{0.12\linewidth}|>{\centering\arraybackslash}p{0.12\linewidth}|}\hline
         Performances&  CS&  MCS& BSS\\\hline
         Charging spped&  Medium&  High& High\\\hline
         Charging cost&  Low&  High& Medium\\\hline
         Distribution density&  High&  Low& Low\\\hline
         Construction cost&  High&  Low& High\\\hline
         Power storage capacity&  Low&  Medium& High\\\hline
 Vehicle compatibility  level& High& Medium&Low\\\hline
 Impacts on grid operation& High& Low&Medium\\ \hline
    \end{tabular}
    
\end{table}

\subsection{Ancillary services} 
The ancillary services contain frequency regulation, spinning reserve, active power support, and reactive power compensation \cite{tan2016integration}. \cite{rahman2022comprehensive,tan2016integration, guille2009conceptual,kisacikoglu2010effects,rotering2010optimal,sortomme2011optimal,ehsani2012vehicle} specifically discuss the implementation methods of V2G ancillary services. Ancillary services can help maintain the balance between generations and loads to ensure stable and reliable power systems. In addition, V2G services also help reduce emissions, increase profits, and provide additional income for EV owners. Meanwhile, using EVs to provide ancillary services presents several challenges. Beyond the issue of battery degradation from frequent bidirectional V2G operations, there is also the need for additional investment in bidirectional chargers. Focusing on the issue of V2G, the automotive energy management system (EMS) deserves more attention due to the ability to coordinate the energy status of EVs under various conditions, which is discussed in Section 4.4. 
 
\subsubsection{Frequency regulation} 
The stability criterion of grid frequency is the power generation must match the load consumption, otherwise the frequency deviation from the criterion operating point will be caused \cite{peng2017dispatching}. \cite{bu2019generic} proposes a generic framework to tackle various frequency-related uncertainties and accommodate different system frequency response models. Given fast response time and low utilization time, EVs with V2G technology are suitable for providing frequency regulation services to the power system \cite{alfaverh2023optimal}. Considering the three layers of frequency control, EVs can perform primary, secondary, and tertiary frequency regulation based on the generator droop characteristic simulation, area control error and economic dispatch, respectively \cite{xiao2013review}. The discharging power of individual EVs is relatively limited, thus EVs usually need to be aggregated and coordinated by an EV aggregator to participate effectively in frequency regulation. Specifically, regulation resources can participate in load frequency control (LFC) through the ancillary service market \cite{cai2022optimal}. In the LFC market, regulation resources submit bids for regulation services in each market time slot, and the system operator allocates LFC regulation capacity according to the market-clearing results. Although LFC regulation signals are issued at a very fast time scale, typically from one second to several seconds, the offered regulation capacity is usually submitted at a much longer time interval, such as on an hourly basis in the PJM Interconnection. When the EV aggregator receives the LFC signal, it dispatches the signal into every single EV so that every regulation resource receive a control signal from the system operator. Dispatching control methods generally contains dispatching based on specifically designed rules, or a pro-rata basis by participation factors. While rule-based method provides control flexibility, it is limited by complex control structures and sampling rate of the frequency regulation signal. Regulation signal intervals of 5–15 minutes \cite{cai2022optimal} cannot fully exploit the advantage of fast response. To overcome these limitations, \cite{cai2022optimal} designs a dispatching control for LFC participation with a faster dispatching time-step to maximize the EV aggregator’s revenue, which is formulated as a nonlinear and nonconvex optimization problem and solved by a genetic algorithm. Simulation results demonstrate that the proposed approach enhances dispatching efficiency, increases regulation capacity, and yields a 6\% revenue improvement. \cite{alfaverh2023optimal} proposes an a DRL-based optimal V2G control strategy to simultaneously maximize the benefits of EV owners and aggregators while performing frequency regulation tasks, where a deep deterministic policy gradient agent is used to dynamically adjust the V2G power scheduling. \cite{jie2023contribution} investigates the contribution of aggregated EV groups to system frequency regulation in a power grid with high PV penetration. Time-series simulations reveal that frequency deviation is reduced to 0\% compared with the initial frequency of 3.5\%.

\subsubsection{Active power support and reactive power compensation}  

Active power support is a service to optimize load curves in the distribution system, which flattens the peak load profile by “peak load shaving” and “load leveling”. Based on bidirectional V2G technologies, excessive EV battery energy is discharged to shave off the peak load and alleviate the applied stress on the distribution system. Furthermore, this service also reduces overall power losses and additional equipment upgrade costs by shaving peak load and maintaining a lower power level in the distribution system \cite{birnie2009solar,rotering2010optimal,zhang2012integration}.

\begin{figure*}[h]  
    \centering      \includegraphics[width=1.0\textwidth]{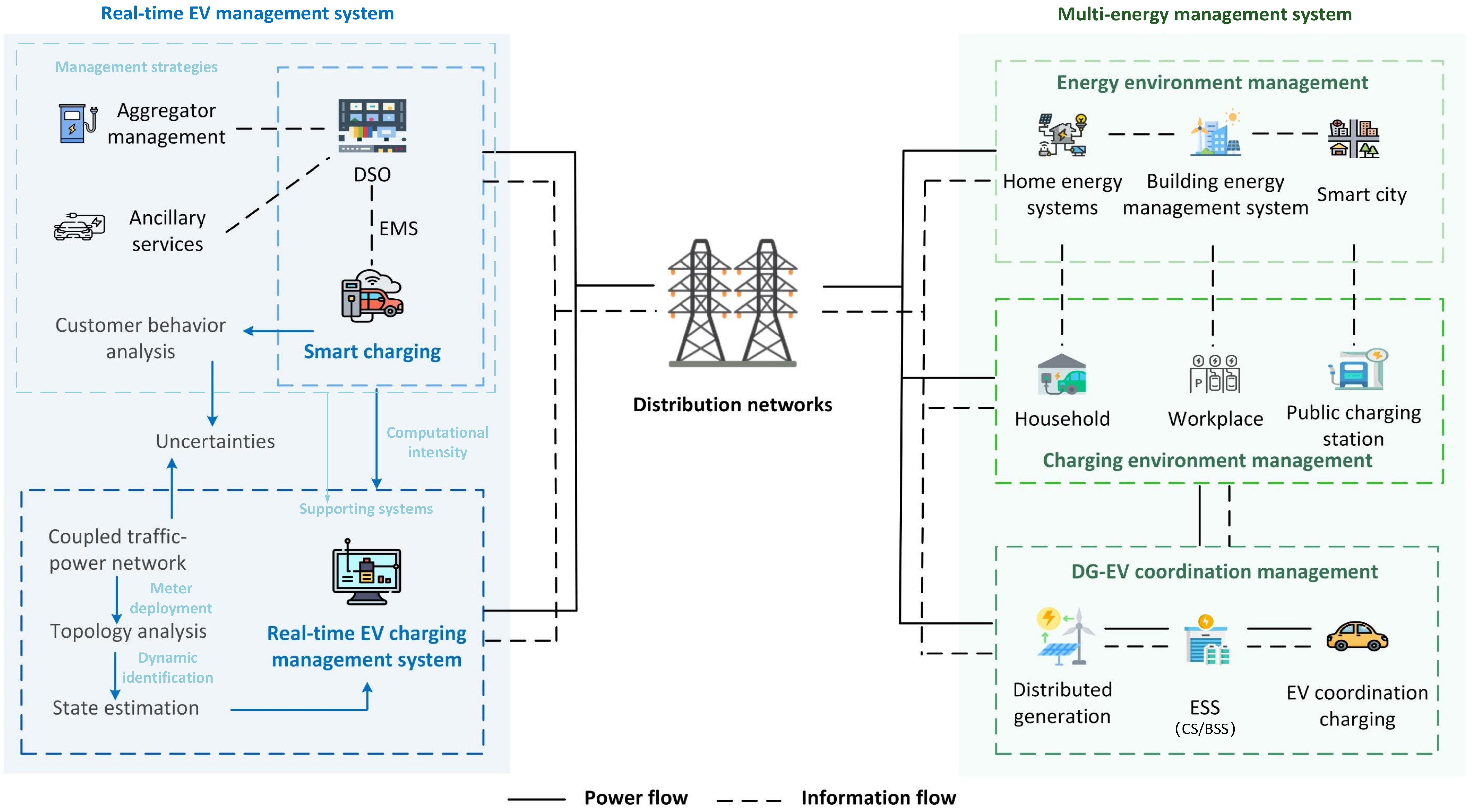}  
    \caption{Management methods to mitigate negative impacts on distribution systems.} 
    \label{fig:3}  
\end{figure*}

Reactive power compensation is a technique to provide voltage regulation and power factor correction \cite{tan2016integration}, further increasing power system operating efficiency and reducing power loss, such as \cite{guille2009conceptual,kisacikoglu2010effects}. In conventional reactive power supply methods, a capacitive reactive power is needed for this function. Distribution generators and static volt–ampere reactive compensators are most commonly used. Based on the capacitive reactive power reserved in the DC-link capacitor of the EV bidirectional battery charger, the EV can supply reactive power compensation by controlling the AC/DC converter without any battery degradation. Management methods to mitigate negative impacts on distribution systems are presented in Figure \ref{fig:3}.  

\section{Real-time EV-grid integration management systems}

\subsection{State estimation} 

The power system is a complex, large-scale, cyber-physical-social system where numerous components interact mutually to maintain a dynamic balance. Integration of EVs will disrupt the current balance, necessitating the establishment of a new equilibrium. As EVs become more prevalent, various parts of the distribution networks will react differently, leading to diverse outcomes and potential impacts. For instance, \cite{sayed2022electric} tests different locations of connecting EVs and demonstrates that EV loads may cause line trips, cascading failures, multiple lines’ overloading, and even a blackout. Based on the topology of distribution networks, real-time state estimation can be conducted to assess the sensitivity of holding EV charging at different locations, and then weaknesses in the network can be identified. It is significant to further enhance monitoring and management of weaknesses during real-time operation to ensure a secure and stable distribution system. The general model of state estimation is written as:
\begin{eqnarray}
    \label{e2}
    \begin{gathered}
Z=h(X,Y)+N,    
    \end{gathered}
\end{eqnarray} 

    \noindent where $Z$ is the measurement vector of measurement, $X$ and $Y$ are vectors of state variables, $N$ is measurement noise, and $h$ is a function that relates state variables to measurements. Knowing the network topology is a prerequisite for evaluating the impact of EV charging on distribution networks \cite{taheri2019milp}. However, due to the limited deployment of meters and infrequently calibrated line parameters, distribution grid operators have problems with accurate and up-to-date real-time information \cite{park2018exact}. Distribution-level phasor measurement unit (PMU) technology \cite{bu2022stability} has a superior capability of monitoring, analyzing and controlling the real-time distribution systems, providing valid technical support for state estimation. Studies \cite{alturki2018marginal,taheri2019milp,park2018exact} have contributed to estimating grid topology using data measured from smart meters and grid sensors \cite{sayed2022electric}. Topology estimation in majority of the power distribution system is hindered by infrequently calibrated line parameters \cite{park2018exact} and sparsely positioned monitoring devices \cite{deka2019topology}, thus limiting the estimation of the rest of the system. To address this issue,
\cite{taheri2019milp} identifies the underlying grid topology by perturbing power injections and analyzing the corresponding voltage responses. This method follows a linear regression setup with unknown vector of line resistances. \cite{moffat2019unsupervised} proposes a noise-robust technique to estimate effective impedances based on the reduced Laplacian form of the Kron reduced admittance matrix. Nevertheless, methods for dynamic topology updates remain limited, especially when considering EV integration. \cite{nie2016system} proposes an effective system state estimation algorithm based on quasi-Newton method that integrates forecast charging load and predictable base power load. \cite{yao2025state} proposes an EV aggregators rescheduling scheme for real-time congestion relief, based on a robust state estimation incorporating measured data from PMU and smart meters. 

With knowledge of power grid topology, methods of analyzing grid stability, such as PV and QV curves, can be utilized to locate buses or nodes where the impact of EV charging can be amplified, thus achieving the identification of weaknesses in the network. Voltage stability indices can be utilized to solve this issue, which have two main classifications \cite{modarresi2016comprehensive}: 1) Jacobian matrix and system variables based VSIs.
2) Bus, line, and overall VSIs.
The voltage stability index for the bus is given by
\begin{eqnarray}
    \label{e2}
    \begin{gathered}
VSI_{i}=\left[1+\left(\frac{I_{i}}{V_{i}}\right)\left(\frac{\Delta V_{i}}{\Delta I_{i}}\right)\right]^{\alpha}, 
    \end{gathered}
\end{eqnarray} 
\noindent where  $I_{i}$  and  $V_{i}$  are the current and voltage at bus $i$ respectively, $\Delta I_{i}$  and  $\Delta V_{i}$ are current and voltage deviation at bus $i$ respectively, and  $\alpha$  is a constant number equal or greater than 1. For instance, the significance of EV charging location is presented by considering different feeders and load curves in \cite{rahman2022comprehensive}. Specifically, \cite{alturki2018marginal} determines the more impactful nodes in assessing network capacity by identifying the maximum possible change in each specific node. \cite{sayed2022electric} simulates the impacts from increased loads on different buses, and obtains the result that buses 6 and 7 are less sensitive with smaller variations than other buses. \cite{deb2018impact} tests the different buses' holding capacity of EV charging by six different cases of EV charging placement and proposes a novel voltage index. It is demonstrated that the system has the capability to support the placement of fast charging stations at strong buses within a certain threshold. However, installing fast charging stations at weak buses adversely affects the stable operation of the power system. \cite{rahman2022comprehensive} compares and analyzes the impact of EV charging loads on the voltage profiles of radial and parallel feeders, resulting in voltage drop across the entire line and a V-curve shape voltage profile. \cite{cheng2014evaluating} proposes a method for extracting the geographical dispersion of EV integrations and evaluates the reliability at load points. \cite{shepero2018modeling} focuses on weak points in city-scale by spatio-temporal modeling of PV and EV. Based on voltage distribution and penetration rates, \cite{ucer2018learning} obtains the conclusion that the congestion condition in the distribution grid can be judged at each local node only by checking its own voltage level. In addition, the issue of attack also deserves consideration; attackers with knowledge of the grid topology can craft smarter attacks by locating the weakest buses that are more likely to cause larger disturbances once attacked, to disrupt the system with smaller numbers of compromised EVs \cite{sayed2022electric}. 

\subsection{System frameworks}

The charging and discharging patterns of EVs are largely influenced by the unpredictable travel habits of their users. Real-time EV management system aims to dynamically generate charging schedules in response to time-varying charging demands and electricity prices \cite{wan2018model}. As EVs are integrated into the distribution system in an unpredictable way, significant uncertainties arise in the timing and location of grid loads. This unpredictability complicates load forecasting and increases the challenges of real-time grid monitoring and management. To address these issues, optimal charging strategies are often designed to maximize the use of renewable energy sources to meet the fluctuating demands of EVs. However, renewable energy generation is intermittent and variable, further aggravating uncertainties in real-time management. To address these issues, many studies have focused on handling the uncertainties of EV charging, renewable energy generation, real-time electricity price, and user behavior, where methods of stochastic optimization \cite{kaewdornhan2023real}, scenario-based stochastic programming \cite{yan2023real,eddahech2012real,yao2016real}, robust optimization \cite{zhao2015risk,vaya2015self}, and dynamic programming \cite{mohamed2013real,you2016real,binetti2015scalable,liao2015dispatch} have been widely applied. Though these approaches can effectively mitigate the impact of uncertainties, many face the challenge of computational dimensionality. For instance, capturing the characteristics of uncertainty requires numerous scenarios, which leads to a heavy computational burden. Even at a moderate scale, the stochastic programming for EV charging dispatching needs several hours to solve \cite{yan2023real}. Many studies focus on addressing the balance between mitigating the impact of uncertainties and high computational intensity. \cite{wu2017two} proposes two control algorithms, SPLET and SAA\_SPLET, to reduce the number of scenarios and alleviate the computational burden. \cite{li2023constrained} introduces a dynamic energy boundary model for EVs to meet charging demand without penalty terms. \cite{wang2019reinforcement} designes four feature functions to approximate the state-action function, reducing the dimension of the state space.  

To address the problem of high computational intensity, decentralized and distributed methods have received considerable attention \cite{chokkalingam2017real,csengor2020real,li2023constrained}. Compared to centralized schemes, decentralized approaches provide better scalability and real-time performance \cite{li2023constrained}. \cite{rivera2014distributed} proposes a distributed anytime algorithm for network utility maximization in real-time EV charging control. This algorithm can operate asynchronously without performance loss and achieve millisecond-level implementation, thereby improving robustness under fast dynamic conditions.  The feasibility of results produced at each iteration ensures the response speed requirement. \cite{latifi2018agent} develops a fully decentralized game-theoretic model to ensure both local optimality for individual EV  and global optimality for the system aggregator. An arbitrary-private topology offers scalability while preserving customer privacy and resilience to node failures. Based on the simulations in a smart microgrid including one aggregator with ten customers, the proposed approach converges fast with one iteration per customer, and reduces grid payment compared with the unscheduled program. \cite{zhou2020admm} designs a hierarchical algorithm based on the alternating direction method of multipliers to allow each individual EV to update its own charging strategy simultaneously, where a decentralized algorithm based on the gradient projection method is further employed to handle the non-separable load regulation term. Compared with the centralized method tested in a 5-feeder and a 12-feeder system with 350 EVs, the average time per outer iteration of the decentralized algorithm is 1.1s and the total computation time is 78s, which is much less than the 2057s required by the centralized method. \cite{binetti2015scalable} proposes a scalable real-time greedy algorithm in a decentralized fashion, which departs from  heavy computations and extensive bi-directional communications. Simulation results confirm its effectiveness under high EV penetration levels. \cite{li2023constrained} proposes a decentralized
framework based on recurrent deep deterministic policy gradient (RDDPG)  and a dynamic energy boundary model of individual EV to achieve discreteness of charging demands, which needs to train only a single agent model to enable local decision-making for multiple charging piles within a charging station.  A RDDPG-based 10-pile-model is tested with a cost reduction of 39.05\% compared with the uncontrolled strategy. The proposed method demonstrates strong scalability and applicability to large-scale EVs without retraining.  To validate this property, the 10-pile model is further extended to scenarios with 20, 30, 40, 50, and 100 piles, confirming the effectiveness under large-scale conditions.

However, the aforementioned distributed methods have limited capability in addressing system-level objectives and overall cooperation requirements. In this regard, the two-level hierarchical control framework has proven effective in improving scalability and separating concerns.  The upper-level controller manages an aggregated representation of EVs and remains agnostic to individual charging behaviors, while the lower-level controller considers detailed models and the specific requirements of each EV \cite{engel2022hierarchical}. However, the main drawback of this approach lies in its limited real-time responsiveness. To address this issue, predictive strategies that combine long-term scheduling with short-term operational control have been applied \cite{latifi2018agent, wu2017two}, striking a balance between the computational burden of large-scale scenarios and the need for real-time decision-making.  In essence, EV charging is a stochastic process influenced by multiple uncertainties that are difficult to quantify. Consequently, real-time management models that decouple from prediction models have attracted significant attention. Representative approaches include model-free reinforcement learning, Lyapunov optimization, and greedy algorithms. The greedy algorithm decomposes the offline problem into subproblems for each time slot, but it lacks guarantees of global optimality.  Lyapunov optimization offers theoretical performance guarantees but is limited in handling highly complex state spaces. Model-free reinforcement learning requires substantial training but demonstrates strong adaptability in dynamic environments. The issue of EV charging dispatching is essentially a time arrangement for EV charging control while considering various uncertainties \cite{li2023constrained}. The current control strategy will influence the system's next state, which increases computational complexity and affects the convergence speed \cite{shen2021real}. Therefore, formulating EV charging control issues as a Markov decision process is a practical approach, where model-free deep reinforcement learning (DRL) techniques can be used as solutions for sequential decision-making problems \cite{zhang2018review} to find optimal strategies in complex and uncertain environments.  \cite{wang2024two} proposes a two-timescale Markov game to enable a model-free and decentralized EV charging control. An action-persistence multi-agent soft actor-critic algorithm is designed to ensure stationary by addressing the hybrid action space of remotely controlled switches and soft open points. \cite{bi2023real} develops a DRL framework for real-time charging decisions, where an off-policy algorithm is designed for offline agent training. Since pre-trained models often fail when confronted with new EVs, changing grids, or evolving user behavior, \cite{zhang2025bi} builds adaptive systems through lifelong or transfer learning, which enables rapid adaptation without retraining from scratch. 

Completing only the performance comparison among different algorithms has considerable limitations for the practical implementation of real-time management strategies. Some studies have considered the influence of management frameworks. For example, in \cite{cai2025online}, the hierarchical scheduling framework reduces computation time by 24.81\%–39.73\% compared with centralized frameworks as the number of EVs increases. The test results in \cite{li2023constrained} show that a decentralized online control algorithm can reduce the total cost by 35.5\%–42.95\%, while accurate day-ahead scheduling can achieve a cost reduction of 50.42\%. Existing studies still lack a comprehensive comparative analysis of centralized, decentralized, hierarchical, and predictive frameworks in terms of real-time control performance. One of the main limitations is the lack of comprehensive charging and discharging behavior datasets and unified benchmarks. In general, decentralized frameworks offer advantages in terms of system scalability, but they are less effective in global coordination and controllability. Hierarchical frameworks can alleviate data pressure, while they may be less competitive in terms of response speed. Predictive frameworks have advantages in processing efficiency and proactive decision-making, but their performance is strongly affected by the accuracy of the prediction module. Moreover, their capability to handle the complexity of real-world operating environments still needs to be further improved.

\subsection{Performance improvement and function expansion}

Due to the randomness of traffic conditions, user behavior, real-time electricity price, the arrival and departure time of EVs, as well as battery states and charging demands, are dynamic and challenging to predict accurately, which makes the management of EV charging challenging \cite{wan2018model}. In recent years, many day-ahead scheduling methods \cite{zhao2015risk,vaya2015self,erdinc2014smart} have been used to address this issue. However, they have limited capability in handling more complex real-time scenarios involving time-varying EV charging demand and electricity price. Real-time management systems with high efficiency and accuracy, along with appropriately designed DR mechanisms, are key factors for distribution system operators in managing EV charging and discharging. The DR mechanism has already received extensive attention in \cite{akhavan2017managing,ghavami2014nonlinear,chekired2017decentralized}, while there are still mismatches between day-ahead scheduling and real-time demand. Therefore, many studies have focused on enhancing the accuracy and efficiency of real-time management. For instance, in \cite{zhao2015risk}, the state-action function is represented by a linear combination of feature functions to transform the decision in the time-varying action space into four time-invariant constants, solving the difficulty in representing the time-varying state and action space in EV charging. \cite{jiang2019real} formulates the charging optimization problem as a cost minimization problem and proposes an improved binary grey wolf optimizer to improve the convergence speed and optimization accuracy. \cite{long2021efficient} develops an ordinal optimization-based method to search for optimal charging strategies within seconds while providing a performance guarantee. \cite{yang2021dynamic} introduces a stochastic dynamic simulation modeling framework for EV fast-charging real-time management, incorporating a multi-server queueing model and a multinomial logit station choice model based on charging prices, expected wait times, and detour distances. \cite{csengor2020real} uses a linear programming method to propose an energy management model for electric vehicle parking lots based on real-time optimization to maximize load factor. \cite{yan2023real} develops a real-time feedback integrated online algorithm based on Lyapunov optimization, which provides theoretical bounds for maximum charging delays. In addition, optimal pricing at EV charging stations plays an important role in guiding EV charging management. Multi-agent reinforcement learning approaches \cite{qian2021multi, li2026mixed, fu2023electric} can be used to model pricing games and determine the optimal prices of EVCSs. In \cite{fu2023electric}, a multi-agent charging-scheduling control framework based on a non-cooperative game is proposed to dynamically adjust EVCS service prices. \cite{li2026mixed} develops a mixed-game pricing model by incorporating the economic dispatch of the distribution network and users’ charging decision responses. A hierarchical attention mechanism is further designed to approximate the Nash equilibrium of pricing games.
\begin{figure*}[h]  
    \centering      \includegraphics[width=1.0\textwidth]{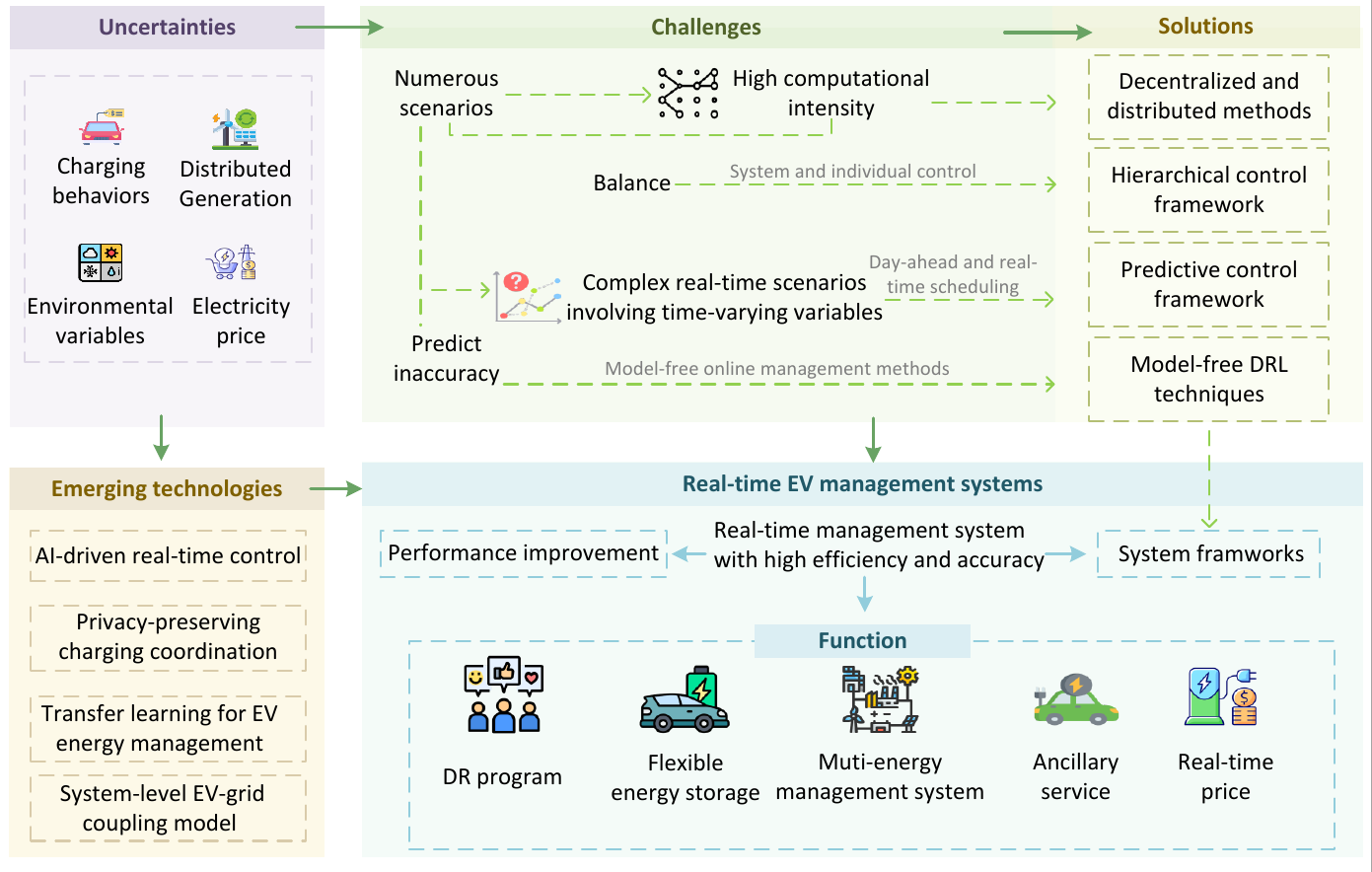}  
    \caption{Framework of real-time EV management systems.}  
    \label{fig:4}  
\end{figure*}

With large-scale EV integration into the power grid and the continuous development of real-time energy management systems, the scope of EV charging has expanded significantly. The coordination of bi-directional energy flows between EV charging systems and other systems, such as photovoltaic, thermal energy, battery swapping, and energy storage systems, has gained widespread attention. Many studies focus on real-time energy management, integrating EV charging with multiple interconnected systems. \cite{zhang2018dynamic} proposes a dynamic cost optimization scheduling method based on real-time information of EV charging demand and PV generation to control the charging process of each EV in a parking lot without considering ESS. \cite{shen2021real} introduced a combined heat and power system and proposed a real-time energy management algorithm for a microgrid based on the Lyapunov optimization technique to minimize the average cost of the microgrid. \cite{jiang2019real} investigates the charging optimization for EVs in parking lots integrated with ESS and PV systems and proposes a real-time EV charging dispatching strategy based on an improved binary grey wolf optimizer. Similar to the management of PEV charging, BSS operation can also be organized through centralized, distributed, and hierarchical management frameworks. \cite{you2017scheduling} investigates centralized solutions based on second-order cone programming relaxation and generalized Benders decomposition, and distributed solutions based on the alternating direction method of multipliers (ADMM) and dual decomposition. In \cite{liu2018distributed}, the optimal operation of a battery swapping-charging system is studied based on a random-permuted ADMM algorithm. In \cite{vsepetanc2019cluster}, a bilevel structure is proposed for BSS operation. The upper level optimizes the charging and discharging schedules of aggregated BSSs, while the lower level provides locational marginal prices for each BSS. This bilevel problem is transformed into a single-level equivalent problem using the Karush-Kuhn-Tucker (KKT) optimality conditions. 

Existing methods still have several limitations. First, many studies lack validation across different datasets, which makes it difficult to demonstrate the general applicability and robustness of the proposed systems. Second, the impacts on power quality are often not comprehensively considered. Third, forecasting errors in load prediction may directly affect the reliability of real-time scheduling decisions. In addition, it remains unclear whether the response speed of real-time dispatch can fully meet the operational requirements of grid management. Other important limitations include communication security risks, information delays, model mismatch, and the ability of management systems to support the rapid growth of EV charging demand while maintaining scalability and generalizability. The summary of the studies on the EV real-time management system is presented in Table \ref{tab3}.

\subsection{Emerging topics}
\subsubsection{Data layer}
Besides the RL methods used for EV management systems mentioned in Section 4.2, many studies have applied AI methods to support other key components of real-time EV management systems, including EV load forecasting, EV user behavior modeling, demand response, and anomaly detection. For example, \cite{singh2024optimizing} proposes an artificial intelligence framework based on predictive demand forecasting and dynamic load distribution, aiming to achieve real-time optimization of EV charging infrastructure. The AI system consists of five key modules: EV charge-demand profiling, EV data augmentation, charge-demand forecasting, forecast explainability, and EV charge optimization. Each module incorporates a series of AI methods, including the k-means algorithm, Gaussian mixture models, multivariate regression, and deterministic optimization \cite{rashid2024comprehensive}.

Many review articles have comprehensively discussed the above fields. For instance, \cite{rashid2024comprehensive} reviews EV charging demand forecasting methods, with a focus on probabilistic and learning-based techniques. \cite{shahriar2020machine} provides a comprehensive review of the applications of supervised and unsupervised machine learning, as well as deep neural networks, in charging behavior analysis and prediction. \cite{wang2024ai} reviews AI-empowered methods for anomaly detection and demand response. Specifically, \cite{yin2024research} adopts partial least squares regression to extract attributes from EV charging load series and uses Bayesian optimization to optimize the hyperparameters of the model, enabling self-adaptive hyperparameter adjustment and reducing the prediction error to 0.092. In \cite{brahmachary2025optimal}, random forest regression is used to predict future load demand, while the 2m-point estimation method is employed to address the uncertainties of EV driving patterns. By combining the transferability of neural networks with the adaptability of DRL to environmental uncertainty, \cite{wang2023transfer} proposes a DRL-based EV charging strategy transfer method to adapt to different charging areas and time periods, increasing the probability of satisfying user charging demands by 47\% and reducing charging costs by 5.4\%. In \cite{sun2026optimizing}, denoising diffusion probabilistic models are used to improve the forecasting accuracy of probabilistic EV loads. \cite{yi2022electric} proposes a time-series forecasting model for charging demand based on the sequence-to-sequence method. This model demonstrates superior performance in multi-step forecasting tests using real-world datasets from the State of Utah and the City of Los Angeles. \cite{khan2025ai} uses a lite transformer -gated recurrent unit, which integrates the transformer attention mechanism with the sequential data-processing advantages of GRU for advanced demand forecasting. By capturing long-term dependencies and prioritizing important features, the model improves forecasting accuracy even under dynamic conditions. In \cite{jin2022deep}, a decentralized decision-making framework based on the deep deterministic policy gradient algorithm is adopted to implement an incentive-based and time-varying demand response mechanism. This framework can balance demand response benefits and user satisfaction while reducing the peak load of charging stations. \cite{wang2023short} develops a series of deep learning methods to predict EV battery swapping demand, including long short-term memory, bidirectional long short-term memory, gated recurrent units, and bidirectional gated recurrent units. In \cite{du2025two}, a hybrid deep learning architecture for electricity demand forecasting in BSSs is proposed by integrating bidirectional temporal convolutional networks with gated recurrent units and attention mechanisms. The model achieves maximum reductions in mean absolute error of 22.87\% and 44.39\% for charging and swapping demand predictions, respectively.

One of the major challenges in EV load prediction is the lack of real-world EV user charging behavior data with spatiotemporal characteristics and multidimensional datasets with sufficient features \cite{rashid2024comprehensive, shahriar2020machine} and multidimensional datasets with sufficient features \cite{shahriar2020machine}. To address these issues, \cite{brinkel2023novel} addresses the lack of forecasting frameworks in the scientific literature that cover all the features required for EV fleet charging scheduling. By transforming charging characteristics into three parameter values describing a virtual battery, the proposed method can predict different virtual battery parameters with high accuracy. \cite{orzechowski2023data} uses weather and other features at public charging stations to predict the demand of multiple charging stations and the entire network, and is the first study to propose machine learning for forecasting medium-term public EV charging demand. Based on data from 11 sites over three years in Scotland, UK, the symmetric mean absolute percentage error for network-level demand forecasting is 5.9\%, and the mean absolute error is 124.7 kWh, which is less than 12\% of the average daily demand.

Although probabilistic-method-based synthetic data generation models have the potential to overcome dataset limitations, further research is still needed to improve the accuracy of EV charging demand forecasting due to their excessive reliance on stochastic behaviors \cite{rashid2024comprehensive}. Compared with a single forecasting model, the combination of different machine learning models can improve forecasting accuracy by addressing issues such as high model complexity, overfitting, and noisy data \cite{rashid2024comprehensive}. For example, deep learning methods can be combined with reinforcement learning to enhance their ability to interact with the environment. In \cite{wang2023transfer}, a transfer method for EV charging strategies based on DRL is proposed to address the problem that charging strategies trained using RL methods may not be applicable across different charging regions and time periods. This method significantly shortens the development cycle of EV charging strategies for different environments. In \cite{khan2025ai}, a game-theory-based DRL method is adopted, considering variables such as vehicle demand, battery capacity, charging speed, and weather conditions, while improving the adaptability of real-time management systems to increasing future demand. In addition, it is necessary to establish standardized guidelines as benchmarks for evaluating model performance, so as to unify different datasets, technical methods, and evaluation metrics \cite{rashid2024comprehensive}.
\subsubsection{Communication layer}
Considering that EV supply equipment and its associated infrastructure may cause severe impacts on downstream systems once compromised, substantial efforts have been devoted to developing cybersecurity strategies. When EVs are connected to electric vehicle charging infrastructure (EVCI), data are mutually exchanged through various communication protocols, such as the Open Charge Point Protocol and IEC 61850 \cite{mitikiri2025regression}. Charging ports are the connection points between EVs and EVCI and are also critical vulnerable points in EVCI \cite{mitikiri2025anomaly}. To protect the power grid from attacks, modern security measures are required \cite{kesavan2025anomaly}. A common attack in the controllers of connected EVs is false data injection (FDI), through which attackers can distort actual energy demand and supply data, thereby posing risks to the secure and reliable operation of the power grid \cite{piperigkos2021impact}. Many studies have focused on anomaly detection for physical equipment data \cite{kesavan2025anomaly,mitikiri2025anomaly}, anomaly detection for charging data \cite{mitikiri2025regression}, and attack defense \cite{yang2022research}. In \cite{kesavan2025anomaly}, a machine-learning-based anomaly detection and Grid Sentinel framework is proposed to safeguard EV charging stations from intrusion, achieving an anomaly detection accuracy of 96.8\%. In \cite{mitikiri2025regression}, a ridge-regression-based machine learning model is adopted to predict SoC changes. Tests using three different spoofing techniques verify the effectiveness of the proposed method. In \cite{mitikiri2025anomaly}, a long short-term memory-based autoencoder model is used to predict the current magnitude at charging ports. The model achieves an accuracy of 98.5\% in predicting current magnitude and identifying anomalies. \cite{yang2022research} proposes a coupled traffic-power framework with a defense mechanism. Within the information network layer of the cyber-physical power system, an active defense model based on static Bayesian game theory is constructed. With the emergence of new types of attacks, several issues deserve further investigation, including the development of more comprehensive datasets, the establishment of early-warning models based on behavioral characteristics rather than specific attack features, and the scalability and robustness of detection models \cite{bean2026cybersecurity}. In addition, privacy-preserving strategies for data exchanged during interactions between EVs and charging facilities also deserve attention \cite{liu2024privacy}. In \cite{liu2024privacy}, a privacy-preserving strategy integrating secure multi-party computation and fully homomorphic encryption is proposed to ensure EV information security. \cite{singh2024optimizing} employs blockchain technology to achieve decentralized and secure communication. \cite{hu2019collaborative} adopts a blockchain consensus mechanism and hash functions to ensure that data are tamper-resistant and traceable. Similarly, \cite{chen2020blockchain} proposes an EV incentive system to maximize renewable energy utilization, in which blockchain technology is used to ensure secure, anonymous, and decentralized operation. In \cite{kong2024privacy}, an improved federated learning method based on local differential privacy is proposed for charging prediction. Tests on a real-world EV fleet dataset demonstrate its adaptability for industrial applications.

\subsubsection{Physical layer}

Furthermore, with the continuous development of vehicle applications, several emerging research directions focusing on physical systems have attracted attention in the field of vehicle-grid interaction, such as grid-forming (GFM) converter interactions, EV optimal design considering grid integration, transfer/lifelong learning for EV energy management and user charging behavior in coupled traffic-power systems. For instance, in \cite{ordono2024grid}, a novel GFM controller for V2G applications is proposed. This controller can ensure stable voltage-source operation and provide grid support regardless of the battery SoC and charging demand. In \cite{barsali2025grid}, the application of GFM control is investigated through detailed simulations of five representative cases. The results show that GFM converters can support islanded grids and black-start procedures and enhance grid resilience against faults.

Focusing on V2G applications, automotive energy management systems deserve more attention because of their ability to coordinate the energy status of EVs under various operating conditions. For instance, \cite{zhang2025scenario} proposes an integrated energy management strategy based on a multi-task DRL algorithm, which dynamically adjusts the reserved SoC to optimize V2G participation while accounting for battery aging costs. Recognizing that reinforcement learning strategies often perform well in simulation but face challenges in real-world deployment, \cite{lei2025sim} establishes an RL-based EMS development toolchain that leverages high-fidelity vehicle models for agent training. \cite{zhang2023integrated} introduces a model-free, multi-state DRL algorithm for integrated thermal and energy management under cold-climate conditions, achieving fuel economy performance close to that of dynamic programming, with a margin of 93.7\%. Traditional design methods usually assume that the training environment and the application environment are relatively similar. However, practical application environments differ significantly and change continuously over time, which may lead to degraded model performance in real-world scenarios. To address this issue, \cite{lian2020cross} proposes a transfer-learning-based method to achieve cross-type knowledge transfer among DRL-based EMSs. Knowledge transfer among four different hybrid electric vehicles (HEVs) is investigated, and the results show that this method can transfer knowledge between two HEVs with substantial structural differences. In \cite{zhang2025bi}, a cross-platform transferable and online-adaptive EMS based on model-agnostic meta-learning is adopted, which can improve fuel economy by 8.0\%-9.5\%, effectively bridging the gap between pre-trained policies and real-world optimal energy management. This framework validates the optimality and knowledge transfer capability of the EMS through real-vehicle experiments. In addition, user charging behavior under complex environmental factors should also be further considered, including the coupled stochastic effects of renewable energy generation and user charging preferences \cite{chen2020blockchain,chung2020intelligent}, interaction management between users and charging facilities \cite{liu2021optimal,he2025unlocking}, and user charging behavior decision-making in coupled traffic-grid systems under complex environmental conditions \cite{mobarak2020solar,fan2026ensuring}. Figure \ref{fig:4} presents the framework of real-time EV management systems. 

\begin{table*}[]
\caption{Summary of the studies on real-time EV-grid management system.} 
\begin{tabular}{>{\centering\arraybackslash}p{0.05\linewidth}>{\centering\arraybackslash}p{0.05\linewidth}>{\raggedright\arraybackslash}p{0.2\linewidth}>{\raggedright\arraybackslash}p{0.4\linewidth}>{\raggedright\arraybackslash}p{0.22\linewidth}}
\hline
 Reference&
  Year &
  Technical foundation &
  Implementation approach &
  Application \\ \hline
\cite{yang2021dynamic} &
  2021 &
  Multi-server queueing model and   EV user behavior model &
  Develop a stochastic dynamic   simulation modeling framework of   EV fast-charging stations &
  Real-time management and   strategic planning \\
\cite{wan2018model} &
  2019 &
  Model-free DRL &
  Extract discriminative features from the electricity prices and approximate the optimal action-value function through a Q network. &
  Optimal strategy for real-time EV charging scheduling \\
\cite{jiang2019real} &
  2019 &
  Improved binary grey wolf   optimizer &
  Short-term PV power prediction and IBGWO &
  Real-time EV charging scheduling strategy \\
\cite{long2021efficient} &
  2021 &
  Aggregated PEV charging model on incomplete Beta function &
  Parameterize the aggregated charging policy using the energy boundaries to express the charging flexibility &
  Real-Time EV charging scheduling \\
\cite{shen2021real} &
  2021 &
  Lyapunov   stochastic optimization &
  Optimize the microgrid average cost without knowing future price, demands, and other system information &
  Real-time energy management algorithm for EV-based microgrid \\
\cite{wang2019reinforcement} &
  2021 &
  Reinforcement learning &
  Develop a model-free data-driven method for EVCSs with random EV arrivals and departures &
  Joint pricing and charging scheduling \\
\cite{csengor2020real} &
  2021 &
  Linear programming &
  Propose a peak load limitation oriented DR program with the objective of maximizing load factor &
  Real-time energy management model for EV parking lot \\
\cite{li2023constrained} &
  2023 &
  Model-free DRL &
  Propose a decentralized framework, a dynamic energy boundary model and formulate a Markov Decision Process &
  Large-scale real-time EV scheduling \\
\cite{chen2021smoothed} &
  2021 &
  Smoothed least-laxity-first algorithm &
  Decide the current charging rates without the knowledge of future arrivals and demands &
  Online decision for EV integration \\
\cite{wan2018data} &
  2018 &
  Data-driven DRL &
  Formulate EV charging management as a Markov Decision Process which has unknown transition probability &
  Optimal EV charging strategy \\
\cite{manbachi2016impact} &
  2016 &
  Real-time simulator and   monitoring platform through DNP.3 protocol &
  Introduce an AMI-based VVO engine to minimize grid loss and Volt–VAR control assets costs while maximizing CVR benefit &
  Real-time co-simulation platform \\
\cite{kaewdornhan2023real} &
  2023 &
  Multi-agent DRL &
  Tackle the multi-home energy management problem with EV charging and discharging scheduling considering uncertainties &
  Real-time multi-home energy management with EV charging scheduling \\
\cite{yan2023real} &
  2024 &
  Lyapunov optimization method &
  Propose an offline model with bounds of the aggregate EV power flexibility region, an online algorithm with a theoretical bound for the maximum charging delay and real-time feedback design&
  Real-time feedback-based online aggregate EV power flexibility characterization \\
\cite{yao2016real}&
  2017 &
  Combination of linear programming and modified convex relaxation &
  Coordinate EV charging loads and accommodate DR programs. &
  Real-time EV charging scheme in the parking station \\

\cite{liang2025online}  &
  2025 & 
  KNN algorithm  &
  Develop an online simulation procedure to evaluate remaining costs corresponding to the current state and candidate actions & EV scheduling \\
\cite{cai2025online} &
  2025 &
  Online hierarchical charging scheduling algorithm &
  Achieve the trade-off between the profits of the charging station operators and the charging satisfaction for the users &
  Online EV charging scheduling \\
\cite{zhou2020admm} &
  2021 &
  ADMM &
  Model the EV fast charging problem as an optimization coordination problem subject to coupled feeder capacity constraints in the distribution network &
  Optimal strategy profile for EVs \\

\cite{latifi2018agent} &
  2019 &
  Decentralized game theoretic model &
  Minimize customers’ payments,   maximize grid efficiency, and provide the maximum potential capacity for ancillary services &
  Real-time dynamic pricing model \\
\cite{chung2020intelligent} &
  2021 &
  Stochastic game &
  Study the interactions between the grid and CSs &
  Intelligent charging management \\
\cite{gao2020deep} &
  2020 &
  DRL  &
  Propose a BSS model to optimize real-time charging/discharging power for charging piles &
  Optimal real-time schedule for a BSS\\
 \hline
\end{tabular}  
\label{tab3}
\end{table*}

\section{Conclusion and future outlooks}
\subsection{Summary of conclusion}
This paper focuses on real-time management for EV-grid integration and provides a comprehensive analysis of the relevant research topics. It first elaborates the negative impacts of EV integration on the real-time operation of distribution systems from the perspective of real-time monitoring. To investigate how these negative impacts can be transformed into positive grid-supporting effects, this paper then presents a systematic elaboration of grid management strategies. Subsequently, real-time management systems, as key components of practical grid operation management, are analyzed in depth. The analysis begins with the real-time state estimation that is the underlying enabling technology, and then discusses major system frameworks as well as methods for optimizing system performance and expanding system functions. The integration of several emerging topics in this field is further reviewed. Finally, the limitations of existing studies and future research directions are discussed.

\subsection{Future research outlooks}
\begin{enumerate}[1)]

\item Datasets: There remains a lack of EV charging and discharging behavior datasets with sufficient environmental information and fine-grained temporal resolution. Existing datasets are mostly collected from specific regions and therefore may not fully represent diverse charging behaviors, grid operating conditions, traffic patterns, weather conditions, and electricity market environments. The availability of more comprehensive datasets and unified benchmarks would significantly facilitate the testing, comparative evaluation, and validation of real-time EV management systems.

\item State estimation: Dynamic state estimation and topology identification for distribution networks in the context of EV-grid integration remain unexplored. As fundamental enabling technologies for real-time management systems, they deserve greater attention in future research.

\item Charging management: Cybersecurity and privacy protection in real-time management systems, especially those involving SCADA systems and communication infrastructures, require further investigation. BSS operation models require considering battery heterogeneity. In addition, the coupling between automotive EMS, EV architecture and power system operation deserves more attention. Furthermore, system-level EV design that considers grid integration, such as battery sizing and powertrain architecture design, represents an important research direction to benefit users, the grid, and society.

\item V2G technology: EV-grid connections introduce new vulnerabilities, creating an urgent need for lightweight intrusion detection and privacy-preserving strategies. Coordinating large-scale and heterogeneous EV fleets remains a major technical challenge, which calls for effective real-time control approaches. Moreover, current market mechanisms is difficult to adequately capture the value of V2G flexibility, highlighting the need to develop new pricing and incentive models.

\item Real-time management platforms: A series of technical challenges arising from large-scale charging data and charging behavior uncertainties require further study. These challenges include cross-scale uncertainty coupling, system processing speed, decision response speed, multidimensional load forecasting accuracy, dispatch decision accuracy, platform scalability, and the balance between real-time performance and optimality. Simulation tests should incorporate multiple datasets and more complex practical operating scenarios, such as fault conditions and real-time variations in distributed generation, to demonstrate model adaptability and generalization capability.
\end{enumerate}

Real-time EV-grid integration management systems can be regarded as part of broader power system management systems. Future studies should further strengthen their coupling with power system operation, including coordinated management with other power equipment, collaborative operation with distributed energy resources and energy hubs, and interaction with real-time electricity markets. Future EV-grid systems will require the joint optimization of user, vehicle, energy, and transportation within a unified framework to ensure overall system efficiency and stability. As a bridge between the energy and transportation sectors, future EV management systems may also undertake broader social functions, such as supporting emergency response and promoting regional energy equity.

\printcredits

\bibliographystyle{unsrt}

\bibliography{cas-refs}

@article{fachrizal2021combined,
  title={Combined PV--EV hosting capacity assessment for a residential LV distribution grid with smart EV charging and PV curtailment},
  author={Fachrizal, Reza and Ramadhani, Umar Hanif and Munkhammar, Joakim and Wid{\'e}n, Joakim},
  journal={Sustainable Energy, Grids and Networks},
  volume={26},
  pages={100445},
  year={2021},
  publisher={Elsevier}
}

@article{kapustin2020long,
  title={Long-term electric vehicles outlook and their potential impact on electric grid},
  author={Kapustin, Nikita O and Grushevenko, Dmitry A},
  journal={Energy Policy},
  volume={137},
  pages={111103},
  year={2020},
  publisher={Elsevier}
}

@article{engel2022hierarchical,
  title={Hierarchical economic model predictive control approach for a building energy management system with scenario-driven EV charging},
  author={Engel, Jens and Schmitt, Thomas and Rodemann, Tobias and Adamy, J{\"u}rgen},
  journal={IEEE Transactions on Smart Grid},
  volume={13},
  number={4},
  pages={3082--3093},
  year={2022},
  publisher={IEEE}
}

@article{li2021coordinating,
  title={Coordinating flexible demand response and renewable uncertainties for scheduling of community integrated energy systems with an electric vehicle charging station: A bi-level approach},
  author={Li, Yang and Han, Meng and Yang, Zhen and Li, Guoqing},
  journal={IEEE Transactions on Sustainable Energy},
  volume={12},
  number={4},
  pages={2321--2331},
  year={2021},
  publisher={IEEE}
}

@article{sayed2022electric,
  title={Electric vehicle attack impact on power grid operation},
  author={Sayed, Mohammad Ali and Atallah, Ribal and Assi, Chadi and Debbabi, Mourad},
  journal={International Journal of Electrical Power \& Energy Systems},
  volume={137},
  pages={107784},
  year={2022},
  publisher={Elsevier}
}

@article{rahman2022comprehensive,
  title={Comprehensive review \& impact analysis of integrating projected electric vehicle charging load to the existing low voltage distribution system},
  author={Rahman, Syed and Khan, Irfan Ahmed and Khan, Ashraf Ali and Mallik, Ayan and Nadeem, Muhammad Faisal},
  journal={Renewable and Sustainable Energy Reviews},
  volume={153},
  pages={111756},
  year={2022},
  publisher={Elsevier}
}

@inproceedings{wu2012pev,
  title={PEV-based combined frequency and voltage regulation for smart grid},
  author={Wu, Chenye and Mohsenian-Rad, Hamed and Huang, Jianwei and Jatskevich, Juri},
  booktitle={2012 IEEE PES Innovative Smart Grid Technologies (ISGT)},
  pages={1--6},
  year={2012},
  organization={IEEE}
}

@article{leemput2015reactive,
  title={Reactive power support in residential LV distribution grids through electric vehicle charging},
  author={Leemput, Niels and Geth, Frederik and Van Roy, Juan and B{\"u}scher, Jeroen and Driesen, Johan},
  journal={Sustainable Energy, Grids and Networks},
  volume={3},
  pages={24--35},
  year={2015},
  publisher={Elsevier}
}

@article{sousa2015multi,
  title={A multi-objective optimization of the active and reactive resource scheduling at a distribution level in a smart grid context},
  author={Sousa, Tiago and Morais, Hugo and Vale, Zita and Castro, Rui},
  journal={Energy},
  volume={85},
  pages={236--250},
  year={2015},
  publisher={Elsevier}
}

@inproceedings{li2012impacts,
  title={Impacts of plug-in hybrid electric vehicles charging on distribution grid and smart charging},
  author={Li, Hui-ling and Bai, Xiao-min and Tan, Wen},
  booktitle={2012 IEEE International Conference on Power System Technology (POWERCON)},
  pages={1--5},
  year={2012},
  organization={IEEE}
}

@article{yong2015review,
  title={A review on the state-of-the-art technologies of electric vehicle, its impacts and prospects},
  author={Yong, Jia Ying and Ramachandaramurthy, Vigna K and Tan, Kang Miao and Mithulananthan, Nadarajah},
  journal={Renewable and sustainable energy reviews},
  volume={49},
  pages={365--385},
  year={2015},
  publisher={Elsevier}
}

@inproceedings{richardson2010impact,
  title={Impact assessment of varying penetrations of electric vehicles on low voltage distribution systems},
  author={Richardson, Peter and Flynn, Damian and Keane, Andrew},
  booktitle={IEEE PES general meeting},
  pages={1--6},
  year={2010},
  organization={IEEE}
}

@inproceedings{gatta2016pq,
  title={PQ and hosting capacity issues for EV charging systems penetration in real MV/LV networks},
  author={Gatta, Fabio Massimo and Geri, Alberto and Lamedica, Regina and Maccioni, Marco and Ruvio, Alessandro},
  booktitle={2016 Power Systems Computation Conference (PSCC)},
  pages={1--7},
  year={2016},
  organization={IEEE}
}

@article{hasanien2015adaptive,
  title={An adaptive control strategy for low voltage ride through capability enhancement of grid-connected photovoltaic power plants},
  author={Hasanien, Hany M},
  journal={IEEE Transactions on power systems},
  volume={31},
  number={4},
  pages={3230--3237},
  year={2015},
  publisher={IEEE}
}

@article{dharmakeerthi2014impact,
  title={Impact of electric vehicle fast charging on power system voltage stability},
  author={Dharmakeerthi, CH and Mithulananthan, N and Saha, Tapan Kumar},
  journal={International Journal of Electrical Power \& Energy Systems},
  volume={57},
  pages={241--249},
  year={2014},
  publisher={Elsevier}
}

@inproceedings{shi2012dynamic,
  title={Dynamic impacts of fast-charging stations for electric vehicles on active distribution networks},
  author={Shi, R and Zhang, Xiao-Ping and Kong, DC and Deng, N and Wang, PY},
  booktitle={IEEE PES innovative smart grid technologies},
  pages={1--6},
  year={2012},
  organization={IEEE}
}

@inproceedings{zhou2016assessment,
  title={Assessment of transient stability support for electric vehicle integration},
  author={Zhou, Bowen and Littler, Tim and Meegahapola, Lasantha},
  booktitle={2016 IEEE Power and Energy Society General Meeting (PESGM)},
  pages={1--5},
  year={2016},
  organization={IEEE}
}

@article{wu2011transient,
  title={Transient stability analysis of SMES for smart grid with vehicle-to-grid operation},
  author={Wu, Diyun and Chau, KT and Liu, Chunhua and Gao, Shuang and Li, Fuhua},
  journal={IEEE Transactions on Applied Superconductivity},
  volume={22},
  number={3},
  pages={5701105--5701105},
  year={2011},
  publisher={IEEE}
}

@inproceedings{mckillop2023impact,
  title={Impact of Electric Vehicles on Power Grids During Network Faults},
  author={McKillop, Nathan and Wembridge, Chris and Franklin, Evan and Lord, James},
  booktitle={2023 IEEE International Conference on Energy Technologies for Future Grids (ETFG)},
  pages={1--6},
  year={2023},
  organization={IEEE}
}

@inproceedings{onar2010grid,
  title={Grid interactions and stability analysis of distribution power network with high penetration of plug-in hybrid electric vehicles},
  author={Onar, Omer C and Khaligh, Alireza},
  booktitle={2010 Twenty-Fifth Annual IEEE Applied Power Electronics Conference and Exposition (APEC)},
  pages={1755--1762},
  year={2010},
  organization={IEEE}
}

@inproceedings{katic2019impact,
  title={Impact of V2G operation of electric vehicle chargers on distribution grid during voltage dips},
  author={Kati{\'c}, Vladimir A and Aleksandar, M and Dumni{\'c}, Boris P and Popadi{\'c}, Bane P and others},
  booktitle={IEEE EUROCON 2019-18th International Conference on Smart Technologies},
  pages={1--6},
  year={2019},
  organization={IEEE}
}

@article{baghaee2019anti,
  title={Anti-islanding protection of PV-based microgrids consisting of PHEVs using SVMs},
  author={Baghaee, Hamid Reza and Mlaki{\'c}, Dragan and Nikolovski, Srete and Dragic{\v{c}}vi{\'c}, Tomislav},
  journal={IEEE Transactions on Smart Grid},
  volume={11},
  number={1},
  pages={483--500},
  year={2019},
  publisher={IEEE}
}

@article{shin2012development,
  title={Development of Smart Grid Monitoring System with Anti-Islanding Function for Electric Vehicle Charging},
  author={Shin, Bum-Sik and Lee, Kyung-Jung and Ro, Sunny and Ki, Y and Ahn, H},
  journal={International Journal of Hybrid Information Technology},
  volume={5},
  number={2},
  pages={269--274},
  year={2012}
}

@article{almeida2015electric,
  title={Electric vehicles contribution for frequency control with inertial emulation},
  author={Almeida, PM Rocha and Soares, Filipe Joel and Lopes, JA Pe{\c{c}}as},
  journal={Electric Power Systems Research},
  volume={127},
  pages={141--150},
  year={2015},
  publisher={Elsevier}
}

@article{pandit2024frequency,
  title={Frequency Support From Electric Vehicles for Advancing Renewable Energy Integration},
  author={Pandit, Dilip and Bera, Atri and Nguyen, Tu and Byrne, Raymond and Chalamala, Babu and Pierre, John and Duan, Dongliang and Nguyen, Nga},
  journal={IEEE Transactions on Power Systems},
  year={2024},
  publisher={IEEE}
}

@article{yadav2023low,
  title={Low voltage ride through capability for resilient electrical distribution system integrated with renewable energy resources},
  author={Yadav, Monika and Pal, Nitai and Saini, Devender Kumar},
  journal={Energy Reports},
  volume={9},
  pages={833--858},
  year={2023},
  publisher={Elsevier}
}

@article{falahi2013potential,
  title={Potential power quality benefits of electric vehicles},
  author={Falahi, Milad and Chou, Hung-Ming and Ehsani, Mehrdad and Xie, Le and Butler-Purry, Karen L},
  journal={IEEE Transactions on sustainable energy},
  volume={4},
  number={4},
  pages={1016--1023},
  year={2013},
  publisher={IEEE}
}

@article{dietmannsberger2017simultaneous,
  title={Simultaneous implementation of LVRT capability and anti-islanding detection in three-phase inverters connected to low-voltage grids},
  author={Dietmannsberger, Markus and Grumm, Florian and Schulz, Detlef},
  journal={ieee transactions on energy conversion},
  volume={32},
  number={2},
  pages={505--515},
  year={2017},
  publisher={IEEE}
}

@article{fachrizal2020smart,
  title={Smart charging of electric vehicles considering photovoltaic power production and electricity consumption: A review},
  author={Fachrizal, Reza and Shepero, Mahmoud and van der Meer, Dennis and Munkhammar, Joakim and Wid{\'e}n, Joakim},
  journal={ETransportation},
  volume={4},
  pages={100056},
  year={2020},
  publisher={Elsevier}
}

@article{stiasny2021sensitivity,
  title={Sensitivity analysis of electric vehicle impact on low-voltage distribution grids},
  author={Stiasny, Jochen and Zufferey, Thierry and Pareschi, Giacomo and Toffanin, Damiano and Hug, Gabriela and Boulouchos, Konstantinos},
  journal={Electric Power Systems Research},
  volume={191},
  pages={106696},
  year={2021},
  publisher={Elsevier}
}

@article{weiller2011plug,
  title={Plug-in hybrid electric vehicle impacts on hourly electricity demand in the United States},
  author={Weiller, Claire},
  journal={Energy Policy},
  volume={39},
  number={6},
  pages={3766--3778},
  year={2011},
  publisher={Elsevier}
}

@article{shafiq2020reliability,
  title={Reliability evaluation of composite power systems: Evaluating the impact of full and plug-in hybrid electric vehicles},
  author={Shafiq, Saifullah and Irshad, Usama Bin and Al-Muhaini, Mohammad and Djokic, Sasa Z and Akram, Umer},
  journal={IEEE access},
  volume={8},
  pages={114305--114314},
  year={2020},
  publisher={IEEE}
}

@article{sortomme2010coordinated,
  title={Coordinated charging of plug-in hybrid electric vehicles to minimize distribution system losses},
  author={Sortomme, Eric and Hindi, Mohammad M and MacPherson, SD James and Venkata, SS},
  journal={IEEE transactions on smart grid},
  volume={2},
  number={1},
  pages={198--205},
  year={2010},
  publisher={IEEE}
}

@article{hartmann2011impact,
  title={Impact of different utilization scenarios of electric vehicles on the German grid in 2030},
  author={Hartmann, Niklas and {\"O}zdemir, Enver Doruk},
  journal={Journal of power sources},
  volume={196},
  number={4},
  pages={2311--2318},
  year={2011},
  publisher={Elsevier}
}

@article{mullan2011modelling,
  title={Modelling the impacts of electric vehicle recharging on the Western Australian electricity supply system},
  author={Mullan, Jonathan and Harries, David and Br{\"a}unl, Thomas and Whitely, Stephen},
  journal={Energy policy},
  volume={39},
  number={7},
  pages={4349--4359},
  year={2011},
  publisher={Elsevier}
}

@article{park2013impact,
  title={Impact of electric vehicle penetration-based charging demand on load profile},
  author={Park, Woo-Jae and Song, Kyung-Bin and Park, Jung-Wook},
  journal={Journal of Electrical Engineering and Technology},
  volume={8},
  number={2},
  pages={244--251},
  year={2013},
  publisher={The Korean Institute of Electrical Engineers}
}

@article{mosaddegh2017optimal,
  title={Optimal demand response for distribution feeders with existing smart loads},
  author={Mosaddegh, Abolfazl and Ca{\~n}izares, Claudio A and Bhattacharya, Kankar},
  journal={IEEE Transactions on Smart Grid},
  volume={9},
  number={5},
  pages={5291--5300},
  year={2017},
  publisher={IEEE}
}

@article{solanki2015including,
  title={Including smart loads for optimal demand response in integrated energy management systems for isolated microgrids},
  author={Solanki, Bharatkumar V and Raghurajan, Akash and Bhattacharya, Kankar and Canizares, Claudio A},
  journal={IEEE Transactions on Smart Grid},
  volume={8},
  number={4},
  pages={1739--1748},
  year={2015},
  publisher={IEEE}
}

@article{hafez2016integrating,
  title={Integrating EV charging stations as smart loads for demand response provisions in distribution systems},
  author={Hafez, Omar and Bhattacharya, Kankar},
  journal={IEEE Transactions on Smart Grid},
  volume={9},
  number={2},
  pages={1096--1106},
  year={2016},
  publisher={IEEE}
}

@article{solanki2017sustainable,
  title={A sustainable energy management system for isolated microgrids},
  author={Solanki, Bharatkumar V and Bhattacharya, Kankar and Canizares, Claudio A},
  journal={IEEE Transactions on Sustainable Energy},
  volume={8},
  number={4},
  pages={1507--1517},
  year={2017},
  publisher={IEEE}
}

@article{wi2013electric,
  title={Electric vehicle charging method for smart homes/buildings with a photovoltaic system},
  author={Wi, Young-Min and Lee, Jong-Uk and Joo, Sung-Kwan},
  journal={IEEE Transactions on Consumer Electronics},
  volume={59},
  number={2},
  pages={323--328},
  year={2013},
  publisher={IEEE}
}

@article{van2016energy,
  title={Energy management system with PV power forecast to optimally charge EVs at the workplace},
  author={Van der Meer, Dennis and Mouli, Gautham Ram Chandra and Mouli, Germ{\'a}n Morales-Espa{\~n}a and Elizondo, Laura Ramirez and Bauer, Pavol},
  journal={IEEE transactions on industrial informatics},
  volume={14},
  number={1},
  pages={311--320},
  year={2016},
  publisher={IEEE}
}

@article{kiviluoma2011methodology,
  title={Methodology for modelling plug-in electric vehicles in the power system and cost estimates for a system with either smart or dumb electric vehicles},
  author={Kiviluoma, Juha and Meibom, Peter},
  journal={Energy},
  volume={36},
  number={3},
  pages={1758--1767},
  year={2011},
  publisher={Elsevier}
}

@article{zheng2018online,
  title={Online distributed MPC-based optimal scheduling for EV charging stations in distribution systems},
  author={Zheng, Yu and Song, Yue and Hill, David J and Meng, Ke},
  journal={IEEE transactions on industrial informatics},
  volume={15},
  number={2},
  pages={638--649},
  year={2018},
  publisher={IEEE}
}

@inproceedings{papadopoulos2010predicting,
  title={Predicting electric vehicle impacts on residential distribution networks with distributed generation},
  author={Papadopoulos, P and Skarvelis-Kazakos, S and Grau, I and Cipcigan, Liana Mirela and Jenkins, N},
  booktitle={2010 IEEE Vehicle Power and Propulsion Conference},
  pages={1--5},
  year={2010},
  organization={IEEE}
}

@article{ramadhani2020review,
  title={Review of probabilistic load flow approaches for power distribution systems with photovoltaic generation and electric vehicle charging},
  author={Ramadhani, Umar Hanif and Shepero, Mahmoud and Munkhammar, Joakim and Wid{\'e}n, Joakim and Etherden, Nicholas},
  journal={International Journal of Electrical Power \& Energy Systems},
  volume={120},
  pages={106003},
  year={2020},
  publisher={Elsevier}
}

@article{liao2015dispatch,
  title={Dispatch of EV charging station energy resources for sustainable mobility},
  author={Liao, Yung-Tang and Lu, Chan-Nan},
  journal={IEEE Transactions on Transportation Electrification},
  volume={1},
  number={1},
  pages={86--93},
  year={2015},
  publisher={IEEE}
}

@article{regulski2014estimation,
  title={Estimation of composite load model parameters using an improved particle swarm optimization method},
  author={Regulski, P and Vilchis-Rodriguez, DS and Djurovi{\'c}, S and Terzija, V},
  journal={IEEE Transactions on Power Delivery},
  volume={30},
  number={2},
  pages={553--560},
  year={2014},
  publisher={IEEE}
}

@article{ju1996nonlinear,
  title={Nonlinear dynamic load modelling: model and parameter estimation},
  author={Ju, Ping and Handschin, E and Karlsson, D},
  journal={IEEE Transactions on Power Systems},
  volume={11},
  number={4},
  pages={1689--1697},
  year={1996},
  publisher={IEEE}
}

@article{lakshminarayanan2018real,
  title={Real-time optimal energy management controller for electric vehicle integration in workplace microgrid},
  author={Lakshminarayanan, Venkateswaran and Chemudupati, Venkata Gowtam S and Pramanick, Sumit Kumar and Rajashekara, Kaushik},
  journal={IEEE Transactions on Transportation Electrification},
  volume={5},
  number={1},
  pages={174--185},
  year={2018},
  publisher={IEEE}
}

@article{ghazvini2017demand,
  title={Demand response implementation in smart households},
  author={Ghazvini, Mohammad Ali Fotouhi and Soares, Jo{\~a}o and Abrishambaf, Omid and Castro, Rui and Vale, Zita},
  journal={Energy and buildings},
  volume={143},
  pages={129--148},
  year={2017},
  publisher={Elsevier}
}

@article{shao2012grid,
  title={Grid integration of electric vehicles and demand response with customer choice},
  author={Shao, Shengnan and Pipattanasomporn, Manisa and Rahman, Saifur},
  journal={IEEE transactions on smart grid},
  volume={3},
  number={1},
  pages={543--550},
  year={2012},
  publisher={IEEE}
}

@article{shepero2018modeling,
  title={Modeling of photovoltaic power generation and electric vehicles charging on city-scale: A review},
  author={Shepero, Mahmoud and Munkhammar, Joakim and Wid{\'e}n, Joakim and Bishop, Justin DK and Bostr{\"o}m, Tobias},
  journal={Renewable and Sustainable Energy Reviews},
  volume={89},
  pages={61--71},
  year={2018},
  publisher={Elsevier}
}

@article{tan2016integration,
  title={Integration of electric vehicles in smart grid: A review on vehicle to grid technologies and optimization techniques},
  author={Tan, Kang Miao and Ramachandaramurthy, Vigna K and Yong, Jia Ying},
  journal={Renewable and Sustainable Energy Reviews},
  volume={53},
  pages={720--732},
  year={2016},
  publisher={Elsevier}
}

@article{guille2009conceptual,
  title={A conceptual framework for the vehicle-to-grid (V2G) implementation},
  author={Guille, Christophe and Gross, George},
  journal={Energy policy},
  volume={37},
  number={11},
  pages={4379--4390},
  year={2009},
  publisher={Elsevier}
}

@inproceedings{kisacikoglu2010effects,
  title={Effects of V2G reactive power compensation on the component selection in an EV or PHEV bidirectional charger},
  author={Kisacikoglu, Mithat C and Ozpineci, Burak and Tolbert, Leon M},
  booktitle={2010 IEEE energy conversion congress and exposition},
  pages={870--876},
  year={2010},
  organization={IEEE}
}

@article{rotering2010optimal,
  title={Optimal charge control of plug-in hybrid electric vehicles in deregulated electricity markets},
  author={Rotering, Niklas and Ilic, Marija},
  journal={IEEE Transactions on Power Systems},
  volume={26},
  number={3},
  pages={1021--1029},
  year={2010},
  publisher={IEEE}
}

@article{sortomme2011optimal,
  title={Optimal scheduling of vehicle-to-grid energy and ancillary services},
  author={Sortomme, Eric and El-Sharkawi, Mohamed A},
  journal={IEEE Transactions on Smart Grid},
  volume={3},
  number={1},
  pages={351--359},
  year={2011},
  publisher={IEEE}
}

@article{ehsani2012vehicle,
  title={Vehicle to grid services: Potential and applications},
  author={Ehsani, Mehrdad and Falahi, Milad and Lotfifard, Saeed},
  journal={Energies},
  volume={5},
  number={10},
  pages={4076--4090},
  year={2012},
  publisher={MDPI}
}

@article{birnie2009solar,
  title={Solar-to-vehicle (S2V) systems for powering commuters of the future},
  author={Birnie III, Dunbar P},
  journal={Journal of Power Sources},
  volume={186},
  number={2},
  pages={539--542},
  year={2009},
  publisher={Elsevier}
}

@article{zhang2012integration,
  title={Integration of PV power into future low-carbon smart electricity systems with EV and HP in Kansai Area, Japan},
  author={Zhang, Qi and Tezuka, Tetsuo and Ishihara, Keiichi N and Mclellan, Benjamin C},
  journal={Renewable Energy},
  volume={44},
  pages={99--108},
  year={2012},
  publisher={Elsevier}
}

@inproceedings{taheri2019milp,
  title={An MILP approach for distribution grid topology identification using inverter probing},
  author={Taheri, Sina and Kekatos, Vassilis and Cavraro, Guido},
  booktitle={2019 IEEE Milan PowerTech},
  pages={1--6},
  year={2019},
  organization={IEEE}
}

@inproceedings{park2018exact,
  title={Exact topology and parameter estimation in distribution grids with minimal observability},
  author={Park, Seiun and Deka, Deepjyoti and Chcrtkov, Michael},
  booktitle={2018 power systems computation conference (PSCC)},
  pages={1--6},
  year={2018},
  organization={IEEE}
}

@inproceedings{alturki2018marginal,
  title={Marginal hosting capacity calculation for electric vehicle integration in active distribution networks},
  author={Alturki, Mansoor and Khodaei, Amin},
  booktitle={2018 IEEE/PES Transmission and Distribution Conference and Exposition (T\&D)},
  pages={1--9},
  year={2018},
  organization={IEEE}
}

@article{cheng2014evaluating,
  title={Evaluating charging service reliability for plug-in EVs from the distribution network aspect},
  author={Cheng, Lin and Chang, Yao and Wu, Qiang and Lin, Weixuan and Singh, Chanan},
  journal={IEEE Transactions on Sustainable Energy},
  volume={5},
  number={4},
  pages={1287--1296},
  year={2014},
  publisher={IEEE}
}

@inproceedings{ucer2018learning,
  title={Learning EV integration impact on a low voltage distribution grid},
  author={Ucer, Emin and Kisacikoglu, Mithat C and Gurbuz, Ali Cafer},
  booktitle={2018 IEEE Power \& Energy Society General Meeting (PESGM)},
  pages={1--5},
  year={2018},
  organization={IEEE}
}

@article{deb2018impact,
  title={Impact of electric vehicle charging station load on distribution network},
  author={Deb, Sanchari and Tammi, Kari and Kalita, Karuna and Mahanta, Pinakeshwar},
  journal={Energies},
  volume={11},
  number={1},
  pages={178},
  year={2018},
  publisher={MDPI}
}

@article{kaewdornhan2023real,
  title={Real-time multi-home energy management with EV charging scheduling using multi-agent deep reinforcement learning optimization},
  author={Kaewdornhan, Niphon and Srithapon, Chitchai and Liemthong, Rittichai and Chatthaworn, Rongrit},
  journal={Energies},
  volume={16},
  number={5},
  pages={2357},
  year={2023},
  publisher={MDPI}
}

@article{yan2023real,
  title={Real-time feedback based online aggregate EV power flexibility characterization},
  author={Yan, Dongxiang and Huang, Shihan and Chen, Yue},
  journal={IEEE Transactions on Sustainable Energy},
  volume={15},
  number={1},
  pages={658--673},
  year={2023},
  publisher={IEEE}
}

@inproceedings{eddahech2012real,
  title={Real-time SOC and SOH estimation for EV Li-ion cell using online parameters identification},
  author={Eddahech, Akram and Briat, Olivier and Vinassa, Jean-Michel},
  booktitle={2012 IEEE Energy Conversion Congress and Exposition (ECCE)},
  pages={4501--4505},
  year={2012},
  organization={IEEE}
}

@article{yao2016real,
  title={A real-time charging scheme for demand response in electric vehicle parking station},
  author={Yao, Leehter and Lim, Wei Hong and Tsai, Teng Shih},
  journal={IEEE Transactions on Smart Grid},
  volume={8},
  number={1},
  pages={52--62},
  year={2016},
  publisher={IEEE}
}

@article{zhao2015risk,
  title={Risk-based day-ahead scheduling of electric vehicle aggregator using information gap decision theory},
  author={Zhao, Jian and Wan, Can and Xu, Zhao and Wang, Jianhui},
  journal={IEEE Transactions on Smart Grid},
  volume={8},
  number={4},
  pages={1609--1618},
  year={2015},
  publisher={IEEE}
}

@article{vaya2015self,
  title={Self scheduling of plug-in electric vehicle aggregator to provide balancing services for wind power},
  author={Vaya, Marina Gonzalez and Andersson, Goeran},
  journal={IEEE Transactions on Sustainable Energy},
  volume={7},
  number={2},
  pages={886--899},
  year={2015},
  publisher={IEEE}
}

@article{mohamed2013real,
  title={Real-time energy management algorithm for plug-in hybrid electric vehicle charging parks involving sustainable energy},
  author={Mohamed, Ahmed and Salehi, Vahid and Ma, Tan and Mohammed, Osama},
  journal={IEEE Transactions on Sustainable Energy},
  volume={5},
  number={2},
  pages={577--586},
  year={2013},
  publisher={IEEE}
}

@article{you2016real,
  title={Real-time state-of-health estimation for electric vehicle batteries: A data-driven approach},
  author={You, Gae-won and Park, Sangdo and Oh, Dukjin},
  journal={Applied energy},
  volume={176},
  pages={92--103},
  year={2016},
  publisher={Elsevier}
}

@article{binetti2015scalable,
  title={Scalable real-time electric vehicles charging with discrete charging rates},
  author={Binetti, Giulio and Davoudi, Ali and Naso, David and Turchiano, Biagio and Lewis, Frank L},
  journal={IEEE Transactions on Smart Grid},
  volume={6},
  number={5},
  pages={2211--2220},
  year={2015},
  publisher={IEEE}
}

@article{yang2021dynamic,
  title={Dynamic modeling and real-time management of a system of EV fast-charging stations},
  author={Yang, Dingtong and Sarma, Navjyoth JS and Hyland, Michael F and Jayakrishnan, R},
  journal={Transportation Research Part C: Emerging Technologies},
  volume={128},
  pages={103186},
  year={2021},
  publisher={Elsevier}
}

@article{wan2018model,
  title={Model-free real-time EV charging scheduling based on deep reinforcement learning},
  author={Wan, Zhiqiang and Li, Hepeng and He, Haibo and Prokhorov, Danil},
  journal={IEEE Transactions on Smart Grid},
  volume={10},
  number={5},
  pages={5246--5257},
  year={2018},
  publisher={IEEE}
}

@article{jiang2019real,
  title={A real-time EV charging scheduling for parking lots with PV system and energy store system},
  author={Jiang, Wei and Zhen, Yongqi},
  journal={IEEE Access},
  volume={7},
  pages={86184--86193},
  year={2019},
  publisher={IEEE}
}

@article{erdinc2014smart,
  title={Smart household operation considering bi-directional EV and ESS utilization by real-time pricing-based DR},
  author={Erdinc, Ozan and Paterakis, Nikolaos G and Mendes, Tiago DP and Bakirtzis, Anastasios G and Catal{\~a}o, Jo{\~a}o PS},
  journal={IEEE Transactions on Smart Grid},
  volume={6},
  number={3},
  pages={1281--1291},
  year={2014},
  publisher={IEEE}
}

@article{long2021efficient,
  title={Efficient real-time EV charging scheduling via ordinal optimization},
  author={Long, Teng and Jia, Qing-Shan and Wang, Gongming and Yang, Yu},
  journal={IEEE Transactions on Smart Grid},
  volume={12},
  number={5},
  pages={4029--4038},
  year={2021},
  publisher={IEEE}
}

@article{shen2021real,
  title={Real-time energy management for microgrid with EV station and CHP generation},
  author={Shen, Zhirong and Wu, Changle and Wang, Lin and Zhang, Guanglin},
  journal={IEEE Transactions on Network Science and Engineering},
  volume={8},
  number={2},
  pages={1492--1501},
  year={2021},
  publisher={IEEE}
}

@article{wang2019reinforcement,
  title={Reinforcement learning for real-time pricing and scheduling control in EV charging stations},
  author={Wang, Shuoyao and Bi, Suzhi and Zhang, Yingjun Angela},
  journal={IEEE Transactions on Industrial Informatics},
  volume={17},
  number={2},
  pages={849--859},
  year={2019},
  publisher={IEEE}
}

@article{chokkalingam2017real,
  title={Real-time forecasting of EV charging station scheduling for smart energy systems},
  author={Chokkalingam, Bharatiraja and Padmanaban, Sanjeevikumar and Siano, Pierluigi and Krishnamoorthy, Ramesh and Selvaraj, Raghu},
  journal={Energies},
  volume={10},
  number={3},
  pages={377},
  year={2017},
  publisher={MDPI}
}

@article{csengor2020real,
  title={Real-time algorithm based intelligent EV parking lot charging management strategy providing PLL type demand response program},
  author={{\c{S}}eng{\"o}r, {\.I}brahim and G{\"u}ner, S{\i}tk{\i} and Erdin{\c{c}}, Ozan},
  journal={IEEE Transactions on Sustainable Energy},
  volume={12},
  number={2},
  pages={1256--1264},
  year={2020},
  publisher={IEEE}
}

@article{li2023constrained,
  title={Constrained large-scale real-time EV scheduling based on recurrent deep reinforcement learning},
  author={Li, Hang and Li, Guojie and Lie, Tek Tjing and Li, Xingzhi and Wang, Keyou and Han, Bei and Xu, Jin},
  journal={International Journal of Electrical Power \& Energy Systems},
  volume={144},
  pages={108603},
  year={2023},
  publisher={Elsevier}
}

@inproceedings{wan2018data,
  title={A data-driven approach for real-time residential EV charging management},
  author={Wan, Zhiqiang and Li, Hepeng and He, Haibo and Prokhorov, Danil},
  booktitle={2018 IEEE Power \& Energy Society General Meeting (PESGM)},
  pages={1--5},
  year={2018},
  organization={IEEE}
}

@inproceedings{rivera2014distributed,
  title={A distributed anytime algorithm for network utility maximization with application to real-time EV charging control},
  author={Rivera, Jose and Jacobsen, Hans-Arno},
  booktitle={53rd IEEE Conference on Decision and Control},
  pages={947--952},
  year={2014},
  organization={IEEE}
}

@article{manbachi2016impact,
  title={Impact of EV penetration on Volt--VAR Optimization of distribution networks using real-time co-simulation monitoring platform},
  author={Manbachi, M and Sadu, A and Farhangi, H and Monti, Antonello and Palizban, Ali and Ponci, Ferdinanda and Arzanpour, Siamak},
  journal={Applied Energy},
  volume={169},
  pages={28--39},
  year={2016},
  publisher={Elsevier}
}

@article{zhou2020admm,
  title={ADMM-based coordination of electric vehicles in constrained distribution networks considering fast charging and degradation},
  author={Zhou, Xu and Zou, Suli and Wang, Peng and Ma, Zhongjing},
  journal={IEEE Transactions on Intelligent Transportation Systems},
  volume={22},
  number={1},
  pages={565--578},
  year={2020},
  publisher={IEEE}
}

@article{akhavan2017managing,
  title={Managing demand for plug-in electric vehicles in unbalanced LV systems with photovoltaics},
  author={Akhavan-Rezai, Elham and Shaaban, Mostafa F and El-Saadany, Ehab F and Karray, Fakhri},
  journal={IEEE Transactions on Industrial Informatics},
  volume={13},
  number={3},
  pages={1057--1067},
  year={2017},
  publisher={IEEE}
}

@inproceedings{ghavami2014nonlinear,
  title={Nonlinear pricing for social optimality of PEV charging under uncertain user preferences},
  author={Ghavami, Abouzar and Kar, Koushik},
  booktitle={2014 48th Annual Conference on Information Sciences and Systems (CISS)},
  pages={1--6},
  year={2014},
  organization={IEEE}
}

@article{chekired2017decentralized,
  title={Decentralized cloud-SDN architecture in smart grid: A dynamic pricing model},
  author={Chekired, Djabir Abdeldjalil and Khoukhi, Lyes and Mouftah, Hussein T},
  journal={IEEE Transactions on Industrial Informatics},
  volume={14},
  number={3},
  pages={1220--1231},
  year={2017},
  publisher={IEEE}
}

@article{zhang2018review,
  title={Review on the research and practice of deep learning and reinforcement learning in smart grids},
  author={Zhang, Dongxia and Han, Xiaoqing and Deng, Chunyu},
  journal={CSEE Journal of Power and Energy Systems},
  volume={4},
  number={3},
  pages={362--370},
  year={2018},
  publisher={CSEE}
}

@article{zhang2018dynamic,
  title={Dynamic charging scheduling for EV parking lots with photovoltaic power system},
  author={Zhang, Yongmin and Cai, Lin},
  journal={IEEE Access},
  volume={6},
  pages={56995--57005},
  year={2018},
  publisher={IEEE}
}

@article{wu2017two,
  title={Two-stage energy management for office buildings with workplace EV charging and renewable energy},
  author={Wu, Di and Zeng, Haibo and Lu, Chao and Boulet, Benoit},
  journal={IEEE Transactions on Transportation Electrification},
  volume={3},
  number={1},
  pages={225--237},
  year={2017},
  publisher={IEEE}
}

@article{latifi2018agent,
  title={Agent-based decentralized optimal charging strategy for plug-in electric vehicles},
  author={Latifi, Milad and Rastegarnia, Amir and Khalili, Azam and Sanei, Saeid},
  journal={IEEE transactions on industrial electronics},
  volume={66},
  number={5},
  pages={3668--3680},
  year={2018},
  publisher={IEEE}
}

@article{rai2019bridgeless,
  title={Bridgeless PFC single ended primary inductance converter in continuous current mode},
  author={Rai, Nor Akmal and Aziz, Mohd Junaidi Abdul and Sahid, Mohd Rodhi and Ayob, Shahrin Md},
  journal={International Journal of Power Electronics and Drive Systems},
  volume={10},
  number={3},
  pages={1427},
  year={2019},
  publisher={IAES Institute of Advanced Engineering and Science}
}

@article{modarresi2016comprehensive,
  title={A comprehensive review of the voltage stability indices},
  author={Modarresi, Javad and Gholipour, Eskandar and Khodabakhshian, Amin},
  journal={Renewable and Sustainable Energy Reviews},
  volume={63},
  pages={1--12},
  year={2016},
  publisher={Elsevier}
}

@article{berg2022data,
  title={A data set of a Norwegian energy community},
  author={Berg, Kjersti and L{\"o}schenbrand, Markus},
  journal={Data in Brief},
  volume={40},
  pages={107683},
  year={2022},
  publisher={Elsevier}
}

@inproceedings{mishra2022intelligent,
  title={An Intelligent Multi-objective EV Charging Architecture with Enhanced Power Quality Operation},
  author={Mishra, Debasish and Singh, Bhim and Panigrahi, BK},
  booktitle={2022 IEEE 2nd International Conference on Sustainable Energy and Future Electric Transportation (SeFeT)},
  pages={1--6},
  year={2022},
  organization={IEEE}
}

@article{alirezaei2016getting,
  title={Getting to net zero energy building: Investigating the role of vehicle to home technology},
  author={Alirezaei, Mehdi and Noori, Mehdi and Tatari, Omer},
  journal={Energy and Buildings},
  volume={130},
  pages={465--476},
  year={2016},
  publisher={Elsevier}
}

@article{liu2021pv,
  title={PV-EV integrated home energy management considering residential occupant behaviors},
  author={Liu, Xuebo and Wu, Yingying and Wu, Hongyu},
  journal={Sustainability},
  volume={13},
  number={24},
  pages={13826},
  year={2021},
  publisher={MDPI}
}

@article{seal2023centralized,
  title={Centralized MPC for home energy management with EV as mobile energy storage unit},
  author={Seal, Sayani and Boulet, Benoit and Dehkordi, Vahid R and Bouffard, Fran{\c{c}}ois and Joos, Geza},
  journal={IEEE Transactions on Sustainable Energy},
  volume={14},
  number={3},
  pages={1425--1435},
  year={2023},
  publisher={IEEE}
}

@inproceedings{garifi2019stochastic,
  title={Stochastic home energy management systems with varying controllable resources},
  author={Garifi, Kaitlyn and Baker, Kyri and Christensen, Dane and Touri, Behrouz},
  booktitle={2019 IEEE Power \& Energy Society General Meeting (PESGM)},
  pages={1--5},
  year={2019},
  organization={IEEE}
}

@inproceedings{irfan2024novel,
  title={A Novel Data-Driven Optimization for Cost-Effective Home Energy Management with PV-EV Integration},
  author={Irfan, Muhammad and Tahir, Tayyab and Deilami, Sara and Huang, Shujuan and Veettil, Binesh Puthen},
  booktitle={2024 IEEE 34th Australasian Universities Power Engineering Conference (AUPEC)},
  pages={1--6},
  year={2024},
  organization={IEEE}
}

@article{hu2019prediction,
  title={The prediction of electric vehicles load profiles considering stochastic charging and discharging behavior and their impact assessment on a real UK distribution network},
  author={Hu, Qian and Li, Haiyu and Bu, Siqi},
  journal={Energy Procedia},
  volume={158},
  pages={6458--6465},
  year={2019},
  publisher={Elsevier}
}

@article{zhang2025scenario,
  title={Scenario-aware electric vehicle energy control with enhanced vehicle-to-grid capability: A multi-task reinforcement learning approach},
  author={Zhang, Hao and Yang, Guixiang and Lei, Nuo and Chen, Chaoyi and Chen, Boli and Qiu, Lin},
  journal={Energy},
  pages={138189},
  year={2025},
  publisher={Elsevier}
}

@article{lei2025sim,
  title={Sim-to-real design and development of reinforcement learning-based energy management strategies for fuel cell electric vehicles},
  author={Lei, Nuo and Zhang, Hao and Hu, Jingjing and Hu, Zunyan and Wang, Zhi},
  journal={Applied Energy},
  volume={393},
  pages={126030},
  year={2025},
  publisher={Elsevier}
}

@article{zhang2023integrated,
  title={Integrated thermal and energy management of connected hybrid electric vehicles using deep reinforcement learning},
  author={Zhang, Hao and Chen, Boli and Lei, Nuo and Li, Bingbing and Li, Rulong and Wang, Zhi},
  journal={IEEE Transactions on Transportation Electrification},
  volume={10},
  number={2},
  pages={4594--4603},
  year={2023},
  publisher={IEEE}
}

@article{peng2017dispatching,
  title={Dispatching strategies of electric vehicles participating in frequency regulation on power grid: A review},
  author={Peng, Chao and Zou, Jianxiao and Lian, Lian},
  journal={Renewable and Sustainable Energy Reviews},
  volume={68},
  pages={147--152},
  year={2017},
  publisher={Elsevier}
}

@article{bu2019generic,
  title={A generic framework for analytical probabilistic assessment of frequency stability in modern power system operational planning},
  author={Bu, Siqi and Wen, Jiaxin and Li, Fangxing},
  journal={IEEE Transactions on Power Systems},
  volume={34},
  number={5},
  pages={3973--3976},
  year={2019},
  publisher={IEEE}
}

@article{alfaverh2023optimal,
  title={Optimal vehicle-to-grid control for supplementary frequency regulation using deep reinforcement learning},
  author={Alfaverh, Fayiz and Dena{\"\i}, Mouloud and Sun, Yichuang},
  journal={Electric Power Systems Research},
  volume={214},
  pages={108949},
  year={2023},
  publisher={Elsevier}
}

@article{cai2022optimal,
  title={Optimal dispatching control of EV aggregators for load frequency control with high efficiency of EV utilization},
  author={Cai, Sinan and Matsuhashi, Ryuji},
  journal={Applied energy},
  volume={319},
  pages={119233},
  year={2022},
  publisher={Elsevier}
}

@article{jie2023contribution,
  title={Contribution to V2G system frequency regulation by charging/discharging control of aggregated EV group},
  author={Jie, Bo and Baba, Jumpei and Kumada, Akiko},
  journal={IEEE Transactions on Industry Applications},
  volume={60},
  number={1},
  pages={1129--1140},
  year={2023},
  publisher={IEEE}
}

@inproceedings{xiao2013review,
  title={Review of the impact of electric vehicles participating in frequency regulation on power grid},
  author={Xiao, Guoxuan and Li, Canbing and Yu, Zhicheng and Cao, Yijia and Fang, Baling},
  booktitle={2013 Chinese automation congress},
  pages={75--80},
  year={2013},
  organization={IEEE}
}

@article{deka2019topology,
  title={Topology estimation using graphical models in multi-phase power distribution grids},
  author={Deka, Deepjyoti and Chertkov, Michael and Backhaus, Scott},
  journal={IEEE Transactions on Power Systems},
  volume={35},
  number={3},
  pages={1663--1673},
  year={2019},
  publisher={IEEE}
}

@article{moffat2019unsupervised,
  title={Unsupervised impedance and topology estimation of distribution networks—Limitations and tools},
  author={Moffat, Keith and Bariya, Mohini and Von Meier, Alexandra},
  journal={IEEE Transactions on Smart Grid},
  volume={11},
  number={1},
  pages={846--856},
  year={2019},
  publisher={IEEE}
}

@article{nie2016system,
  title={System state estimation considering EV penetration with unknown behavior using quasi-Newton method},
  author={Nie, Yongquan and Chung, CY and Xu, NZ},
  journal={IEEE Transactions on Power Systems},
  volume={31},
  number={6},
  pages={4605--4615},
  year={2016},
  publisher={IEEE}
}

@article{yao2025state,
  title={State estimation-aided real time congestion relief using EV aggregators in distribution system},
  author={Yao, Yuan and Xu, Yinliang and Sun, Hongbin},
  journal={CSEE Journal of Power and Energy Systems},
  year={2025},
  publisher={CSEE}
}

@article{zhang2025bi,
  title={Bi-Level Transfer Learning for Lifelong-Intelligent Energy Management of Electric Vehicles},
  author={Zhang, Hao and Lei, Nuo and Peng, Wang and Li, Bingbing and Lv, Shujun and Chen, Boli and Wang, Zhi},
  journal={IEEE Transactions on Intelligent Transportation Systems},
  year={2025},
  publisher={IEEE}
}

@article{wang2024two,
  title={Two-Timescale Voltage Regulation for Coupled Distribution and Flash-Charging-Enabled Public Transit Systems Using Deep Reinforcement Learning},
  author={Wang, Ruoheng and Bi, Xiaowen and Bu, Siqi and Long, Meng},
  journal={IEEE Transactions on Transportation Electrification},
  year={2024},
  publisher={IEEE}
}

@article{bi2023real,
  title={Real-time scheduling of electric bus flash charging at intermediate stops: A deep reinforcement learning approach},
  author={Bi, Xiaowen and Wang, Ruoheng and Ye, Hongbo and Hu, Qian and Bu, Siqi and Chung, Edward},
  journal={IEEE Transactions on Transportation Electrification},
  volume={10},
  number={3},
  pages={6309--6324},
  year={2023},
  publisher={IEEE}
}

@article{bu2022stability,
  title={Stability and dynamics of active distribution networks (ADNs) with D-PMU technology: A review},
  author={Bu, Siqi and Meegahapola, Lasantha Gunaruwan and Wadduwage, Darshana Prasad and Foley, Aoife M},
  journal={IEEE Transactions on Power Systems},
  volume={38},
  number={3},
  pages={2791--2804},
  year={2022},
  publisher={IEEE}
}

@article{dahiwale2024comprehensive,
  title={A comprehensive review of smart charging strategies for electric vehicles and way forward},
  author={Dahiwale, Payal Vyankat and Rather, Zakir H and Mitra, Indradip},
  journal={IEEE Transactions on Intelligent Transportation Systems},
  volume={25},
  number={9},
  pages={10462--10482},
  year={2024},
  publisher={IEEE}
}

@article{wang2022cyber,
  title={A cyber--physical--social perspective on future smart distribution systems},
  author={Wang, Yi and Chen, Chien-Fei and Kong, Peng-Yong and Li, Husheng and Wen, Qingsong},
  journal={Proceedings of the IEEE},
  volume={111},
  number={7},
  pages={694--724},
  year={2022},
  publisher={IEEE}
}

@article{hasan2023distribution,
  title={Distribution network voltage analysis with data-driven electric vehicle load profiles},
  author={Hasan, Kazi N and Muttaqi, Kashem M and Borboa, Pablo and Scira, Jakem and Zhang, Zihao and Leishman, Matthew},
  journal={Sustainable Energy, Grids and Networks},
  volume={36},
  pages={101216},
  year={2023},
  publisher={Elsevier}
}

@article{senol2024harmonics,
  title={Harmonics measurement, analysis, and impact assessment of electric vehicle smart charging},
  author={Senol, Murat and Bayram, I Safak and Hunter, Lewis and Sevdari, Kristian and McGarry, Connor and Gaona, David Campos and Gehrke, Oliver and Galloway, Stuart},
  journal={IEEE Open Journal of Vehicular Technology},
  volume={6},
  pages={109--127},
  year={2024},
  publisher={IEEE}
}

@article{khalid2019comprehensive,
  title={A Comprehensive review on electric vehicles charging infrastructures and their impacts on power-quality of the utility grid},
  author={Khalid, Mohd Rizwan and Alam, Mohammad Saad and Sarwar, Adil and Asghar, MS Jamil},
  journal={ETransportation},
  volume={1},
  pages={100006},
  year={2019},
  publisher={Elsevier}
}

@article{srivastava2023electric,
  title={Electric vehicle integration’s impacts on power quality in distribution network and associated mitigation measures: a review},
  author={Srivastava, Abhinav and Manas, Munish and Dubey, Rajesh Kumar},
  journal={Journal of Engineering and Applied Science},
  volume={70},
  number={1},
  pages={32},
  year={2023},
  publisher={Springer}
}

@article{mitikiri2025regression,
  title={Regression based anomaly detection in electric vehicle state of charge fluctuations through analysis of electric vehicle charging infrastructure data},
  author={Mitikiri, Sagar Babu and Tiwari, Yash and Srinivas, Vedantham Lakshmi and Pal, Mayukha},
  journal={Sustainable Energy, Grids and Networks},
  volume={42},
  pages={101704},
  year={2025},
  publisher={Elsevier}
}

@article{mitikiri2025anomaly,
  title={Anomaly detection of adversarial cyber attacks on electric vehicle charging stations},
  author={Mitikiri, Sagar Babu and Srinivas, Vedantham Lakshmi and Pal, Mayukha},
  journal={e-Prime-Advances in Electrical Engineering, Electronics and Energy},
  volume={11},
  pages={100911},
  year={2025},
  publisher={Elsevier}
}

@article{kesavan2025anomaly,
  title={Anomaly detection with grid sentinel framework for electric vehicle charging stations in a smart grid environment},
  author={Kesavan, V Thiruppathy and Hossen, Md Jakir and Gopi, R and Joseph, Emerson Raja},
  journal={Scientific Reports},
  volume={15},
  number={1},
  pages={15774},
  year={2025},
  publisher={Nature Publishing Group UK London}
}

@article{piperigkos2021impact,
  title={Impact of false data injection attacks on decentralized electric vehicle charging protocols},
  author={Piperigkos, Nikos and Lalos, Aris S},
  journal={Transportation Research Procedia},
  volume={52},
  pages={331--338},
  year={2021},
  publisher={Elsevier}
}

@article{yang2022research,
  title={Research on security defense of coupled transportation and cyber-physical power system based on the static Bayesian game},
  author={Yang, Zian and Xiang, Yueping and Liao, Kai and Yang, Jianwei},
  journal={IEEE Transactions on Intelligent Transportation Systems},
  volume={24},
  number={3},
  pages={3571--3583},
  year={2022},
  publisher={IEEE}
}

@article{bean2026cybersecurity,
  title={Cybersecurity of Electric Vehicle Charging Infrastructure: Recent Advances, Open Challenges, and Future Directions},
  author={Bean, Joshua and Manias, Dimitrios Michael},
  journal={arXiv preprint arXiv:2605.24190},
  year={2026}
}

@article{liu2024privacy,
  title={Privacy-preserving electric vehicle charging recommendation by incorporating full homomorphic encryption and secure multi-party computing},
  author={Liu, Yiqi and Ju, Jiaxin and Li, Zhiyi},
  journal={World Electric Vehicle Journal},
  volume={15},
  number={10},
  pages={446},
  year={2024},
  publisher={MDPI}
}

@article{singh2024optimizing,
  title={Optimizing demand response and load balancing in smart EV charging networks using AI integrated blockchain framework},
  author={Singh, Arvind R and Kumar, R Seshu and Madhavi, K Reddy and Alsaif, Faisal and Bajaj, Mohit and Zaitsev, Ievgen},
  journal={Scientific Reports},
  volume={14},
  number={1},
  pages={31768},
  year={2024},
  publisher={Nature Publishing Group UK London}
}

@article{lian2020cross,
  title={Cross-type transfer for deep reinforcement learning based hybrid electric vehicle energy management},
  author={Lian, Renzong and Tan, Huachun and Peng, Jiankun and Li, Qin and Wu, Yuankai},
  journal={IEEE Transactions on Vehicular Technology},
  volume={69},
  number={8},
  pages={8367--8380},
  year={2020},
  publisher={IEEE}
}

@article{ordono2024grid,
  title={A grid forming controller with integrated state of charge management for V2G chargers},
  author={Ordono, Ander and Asensio, Francisco Javier and Cortajarena, Jose Antonio and Zamora, Inmaculada and Gonz{\'a}lez-P{\'e}rez, Mikel and Salda{\~n}a, Gaizka},
  journal={International Journal of Electrical Power \& Energy Systems},
  volume={157},
  pages={109862},
  year={2024},
  publisher={Elsevier}
}

@article{ibrahim2024analysis,
  title={Analysis of multidimensional impacts of electric vehicles penetration in distribution networks},
  author={Ibrahim, Rania A and Gaber, Ibrahim M and Zakzouk, Nahla E},
  journal={Scientific Reports},
  volume={14},
  number={1},
  pages={27854},
  year={2024},
  publisher={Nature Publishing Group UK London}
}

@article{rana2025comprehensive,
  title={Comprehensive review on the charging technologies of electric vehicles (EV) and their impact on power grid},
  author={Rana, Mohammed Masud and Alam, SM Mahfuz and Rafi, Faiaz Allahma and Deb, Swarup Bashu and Agili, Boker and He, Miao and Ali, Mohd Hasan},
  journal={IEEE Access},
  volume={13},
  pages={35124--35156},
  year={2025},
  publisher={IEEE}
}

@article{inci2024power,
  title={Power system integration of electric vehicles: A review on impacts and contributions to the smart grid},
  author={Inci, Mustafa and {\c{C}}elik, {\"O}zg{\"u}r and Lashab, Abderezak and Bay{\i}nd{\i}r, Kamil {\c{C}}a{\u{g}}atay and Vasquez, Juan C and Guerrero, Josep M},
  journal={Applied Sciences},
  volume={14},
  number={6},
  pages={2246},
  year={2024},
  publisher={MDPI}
}

@article{motlagh2025review,
  title={A review on electric vehicle charging station operation considering market dynamics and grid interaction},
  author={Motlagh, Saheb Ghanbari and Oladigbolu, Jamiu and Li, Li},
  journal={Applied Energy},
  volume={392},
  pages={126058},
  year={2025},
  publisher={Elsevier}
}

@article{suresh2026comprehensive,
  title={A comprehensive review on planning of EV charging infrastructure in PDN: Modeling elastic charging behavior and impact analysis},
  author={Suresh, Tejavath and Shah, Varsha Ajit and Shukla, Akanksha},
  journal={Renewable and Sustainable Energy Reviews},
  volume={239},
  pages={117110},
  year={2026},
  publisher={Elsevier}
}

@article{wang2021grid,
  title={Grid impact of electric vehicle fast charging stations: Trends, standards, issues and mitigation measures-an overview},
  author={Wang, Lu and Qin, Zian and Slangen, Tim and Bauer, Pavol and Van Wijk, Thijs},
  journal={IEEE Open Journal of Power Electronics},
  volume={2},
  pages={56--74},
  year={2021},
  publisher={IEEE}
}

@article{das2020electric,
  title={Electric vehicles standards, charging infrastructure, and impact on grid integration: A technological review},
  author={Das, Himadry Shekhar and Rahman, Mohammad Mominur and Li, Shuhui and Tan, Chee Wei},
  journal={Renewable and Sustainable Energy Reviews},
  volume={120},
  pages={109618},
  year={2020},
  publisher={Elsevier}
}

@article{ronanki2023electric,
  title={Electric vehicle charging infrastructure: Review, cyber security considerations, potential impacts, countermeasures, and future trends},
  author={Ronanki, Deepak and Karneddi, Harish},
  journal={IEEE Journal of Emerging and Selected Topics in Power Electronics},
  volume={12},
  number={1},
  pages={242--256},
  year={2023},
  publisher={IEEE}
}

@article{khalid2021comprehensive,
  title={A comprehensive review on structural topologies, power levels, energy storage systems, and standards for electric vehicle charging stations and their impacts on grid},
  author={Khalid, Mohd Rizwan and Khan, Irfan A and Hameed, Salman and Asghar, M Syed Jamil and Ro, JongSuk},
  journal={IEEE access},
  volume={9},
  pages={128069--128094},
  year={2021},
  publisher={IEEE}
}

@article{ahmed2026holistic,
  title={A holistic review of electric vehicle charging impacts on power distribution networks: Technical challenges, smart mitigation strategies and future directions},
  author={Ahmed, Fahim and Ahmad, Shameem and Rahman, Mir Toufikur and Hazari, Md Rifat and Faiz, Rethwan and Ahmed, Tofael and Karimi, Mazaher},
  journal={Applied Energy},
  volume={402},
  pages={126961},
  year={2026},
  publisher={Elsevier}
}

@article{rashid2024comprehensive,
  title={A comprehensive survey of electric vehicle charging demand forecasting techniques},
  author={Rashid, Mamunur and Elfouly, Tarek and Chen, Nan},
  journal={IEEE Open Journal of Vehicular Technology},
  volume={5},
  pages={1348--1373},
  year={2024},
  publisher={IEEE}
}

@article{wang2024ai,
  title={AI-empowered methods for smart energy consumption: A review of load forecasting, anomaly detection and demand response},
  author={Wang, Xinlin and Wang, Hao and Bhandari, Binayak and Cheng, Leming},
  journal={International Journal of Precision Engineering and Manufacturing-Green Technology},
  volume={11},
  number={3},
  pages={963--993},
  year={2024},
  publisher={Springer}
}

@article{wang2023transfer,
  title={A transfer learning method for electric vehicles charging strategy based on deep reinforcement learning},
  author={Wang, Kang and Wang, Haixin and Yang, Zihao and Feng, Jiawei and Li, Yanzhen and Yang, Junyou and Chen, Zhe},
  journal={Applied Energy},
  volume={343},
  pages={121186},
  year={2023},
  publisher={Elsevier}
}

@article{yin2024research,
  title={Research on EV charging load forecasting and orderly charging scheduling based on model fusion},
  author={Yin, Wanjun and Ji, Jianbo},
  journal={Energy},
  volume={290},
  pages={130126},
  year={2024},
  publisher={Elsevier}
}

@article{brahmachary2025optimal,
  title={Optimal distribution network expansion and EV charging station allocation based on load forecasted using machine learning},
  author={Brahmachary, Rupali and Ahmed, Irfan},
  journal={IEEE Transactions on Industry Applications},
  year={2025},
  publisher={IEEE}
}

@article{sun2026optimizing,
  title={Optimizing Electric Vehicle Charging Load Forecasting via Ensemble of Diffusion Models},
  author={Sun, Xijuan and Wu, Di and Jenkin, Michael and Zinflou, Arnaud and Wang, Boyu and Boulet, Benoit},
  journal={IEEE Transactions on Intelligent Vehicles},
  year={2026},
  publisher={IEEE}
}

@article{yi2022electric,
  title={Electric vehicle charging demand forecasting using deep learning model},
  author={Yi, Zhiyan and Liu, Xiaoyue Cathy and Wei, Ran and Chen, Xi and Dai, Jiangpeng},
  journal={Journal of Intelligent Transportation Systems},
  volume={26},
  number={6},
  pages={690--703},
  year={2022},
  publisher={Taylor \& Francis}
}

@article{orzechowski2023data,
  title={A data-driven framework for medium-term electric vehicle charging demand forecasting},
  author={Orzechowski, Alexander and Lugosch, Loren and Shu, Hao and Yang, Raymond and Li, Wei and Meyer, Brett H},
  journal={Energy and AI},
  volume={14},
  pages={100267},
  year={2023},
  publisher={Elsevier}
}

@article{brinkel2023novel,
  title={A novel forecasting approach to schedule aggregated electric vehicle charging},
  author={Brinkel, Nico and Visser, Lennard and van Sark, Wilfried and AlSkaif, Tarek},
  journal={Energy and AI},
  volume={14},
  pages={100297},
  year={2023},
  publisher={Elsevier}
}

@article{khan2025ai,
  title={AI-Driven Dynamic Allocation and Management Optimization for EV Charging Stations},
  author={Khan, Arfat Ahmad and Mahendran, Rakesh Kumar and Ullah, Fasee and Ali, Farman and Bashir, Ali Kashif and Al Dabel, Maryam M and Omar, Marwan},
  journal={IEEE Transactions on Intelligent Transportation Systems},
  year={2025},
  publisher={IEEE}
}

@article{jin2022deep,
  title={Deep reinforcement learning-based strategy for charging station participating in demand response},
  author={Jin, Ruiyang and Zhou, Yuke and Lu, Chao and Song, Jie},
  journal={Applied Energy},
  volume={328},
  pages={120140},
  year={2022},
  publisher={Elsevier}
}

@article{shahriar2020machine,
  title={Machine learning approaches for EV charging behavior: A review},
  author={Shahriar, Sakib and Al-Ali, Abdul-Rahman and Osman, Ahmed H and Dhou, Salam and Nijim, Mais},
  journal={Ieee Access},
  volume={8},
  pages={168980--168993},
  year={2020},
  publisher={IEEE}
}

@article{barsali2025grid,
  title={Grid forming inverters for electric vehicle charging stations to enhance distribution grid resilience},
  author={Barsali, Stefano and Bojoi, Radu and Ceraolo, Massimo and Mallemaci, Vincenzo and Mandrile, Fabio and Mocci, Susanna and Pasini, Gianluca},
  journal={IEEE Access},
  year={2025},
  publisher={IEEE}
}

@article{cai2025online,
  title={Online Coordination of Electric Vehicle Charging for Balanced Operating Profit and User Dissatisfaction},
  author={Cai, Ling and Guo, Ge and Shi, Lengandong and Ma, Miaomiao},
  journal={IEEE Transactions on Vehicular Technology},
  year={2025},
  publisher={IEEE}
}

@article{chen2021smoothed,
  title={Smoothed least-laxity-first algorithm for electric vehicle charging: Online decision and performance analysis with resource augmentation},
  author={Chen, Niangjun and Kurniawan, Christian and Nakahira, Yorie and Chen, Lijun and Low, Steven H},
  journal={IEEE transactions on smart grid},
  volume={13},
  number={3},
  pages={2209--2217},
  year={2021},
  publisher={IEEE}
}

@article{liang2025online,
  title={Online Operation of Renewable Energy and Battery Integrated Electric Vehicle Parking Lots with an Improved Priority Rule},
  author={Liang, Runze and Wei, Wei and Wang, Zhaojian and Liu, Feng},
  journal={IEEE Transactions on Transportation Electrification},
  year={2025},
  publisher={IEEE}
}

@article{kong2024privacy,
  title={Privacy-preserving estimation of electric vehicle charging behavior: A federated learning approach based on differential privacy},
  author={Kong, Xiuping and Lu, Lin and Xiong, Ke},
  journal={Internet of Things},
  volume={28},
  pages={101344},
  year={2024},
  publisher={Elsevier}
}

@article{zhang2021stochastic,
  title={Stochastic modeling and analysis of public electric vehicle fleet charging station operations},
  author={Zhang, Tianyang and Chen, Xi and Wu, Bin and Dedeoglu, Mehmet and Zhang, Junshan and Trajkovic, Ljiljana},
  journal={IEEE Transactions on Intelligent Transportation Systems},
  volume={23},
  number={7},
  pages={9252--9265},
  year={2021},
  publisher={IEEE}
}

@article{zhang2018monte,
  title={A Monte Carlo simulation approach to evaluate service capacities of EV charging and battery swapping stations},
  author={Zhang, Tianyang and Chen, Xi and Yu, Zhe and Zhu, Xiaoyan and Shi, Di},
  journal={IEEE Transactions on Industrial Informatics},
  volume={14},
  number={9},
  pages={3914--3923},
  year={2018},
  publisher={IEEE}
}

@article{wu2021survey,
  title={A survey of battery swapping stations for electric vehicles: Operation modes and decision scenarios},
  author={Wu, Hao},
  journal={IEEE Transactions on Intelligent Transportation Systems},
  volume={23},
  number={8},
  pages={10163--10185},
  year={2021},
  publisher={IEEE}
}

@article{chen2021electric,
  title={An electric vehicle battery-swapping system: Concept, architectures, and implementations},
  author={Chen, Xi and Xing, Kai and Ni, Feng and Wu, Yujie and Xia, Yongxiang},
  journal={IEEE Intelligent Transportation Systems Magazine},
  volume={14},
  number={5},
  pages={175--194},
  year={2021},
  publisher={IEEE}
}

@article{chen2020blockchain,
  title={Blockchain-based electric vehicle incentive system for renewable energy consumption},
  author={Chen, Xi and Zhang, Tianyang and Ye, Wenxing and Wang, Zhiwei and Iu, Herbert Ho-Ching},
  journal={IEEE Transactions on Circuits and Systems II: Express Briefs},
  volume={68},
  number={1},
  pages={396--400},
  year={2020},
  publisher={IEEE}
}

@article{chung2020intelligent,
  title={Intelligent charging management of electric vehicles considering dynamic user behavior and renewable energy: A stochastic game approach},
  author={Chung, Hwei-Ming and Maharjan, Sabita and Zhang, Yan and Eliassen, Frank},
  journal={IEEE Transactions on Intelligent Transportation Systems},
  volume={22},
  number={12},
  pages={7760--7771},
  year={2020},
  publisher={IEEE}
}

@article{fu2023electric,
  title={Electric vehicle charging scheduling control strategy for the large-scale scenario with non-cooperative game-based multi-agent reinforcement learning},
  author={Fu, Liyue and Wang, Tong and Song, Min and Zhou, Yuhu and Gao, Shan},
  journal={International Journal of Electrical Power \& Energy Systems},
  volume={153},
  pages={109348},
  year={2023},
  publisher={Elsevier}
}

@article{qian2021multi,
  title={Multi-agent deep reinforcement learning method for EV charging station game},
  author={Qian, Tao and Shao, Chengcheng and Li, Xuliang and Wang, Xiuli and Chen, Zhiping and Shahidehpour, Mohammad},
  journal={IEEE Transactions on Power Systems},
  volume={37},
  number={3},
  pages={1682--1694},
  year={2021},
  publisher={IEEE}
}

@article{li2026mixed,
  title={Mixed Competitive—Cooperative Pricing for EV Charging Stations: A Multi-Agent Reinforcement Learning Approach with Heterogeneous Hierarchical Attention},
  author={Li, Yujing and Xing, Qiang and Lv, Si and Li, Zening},
  journal={IEEE Transactions on Smart Grid},
  year={2026},
  publisher={IEEE}
}

@article{you2017scheduling,
  title={Scheduling of EV battery swapping--Part II: Distributed solutions},
  author={You, Pengcheng and Low, Steven H and Zhang, Liang and Deng, Ruilong and Giannakis, Georgios B and Sun, Youxian and Yang, Zaiyue},
  journal={IEEE Transactions on Control of Network Systems},
  volume={5},
  number={4},
  pages={1920--1930},
  year={2017},
  publisher={IEEE}
}

@article{liu2018distributed,
  title={Distributed operation management of battery swapping-charging systems},
  author={Liu, Xiaochuan and Zhao, Tianyang and Yao, Shuhan and Soh, Cheong Boon and Wang, Peng},
  journal={IEEE Transactions on Smart Grid},
  volume={10},
  number={5},
  pages={5320--5333},
  year={2018},
  publisher={IEEE}
}

@article{vsepetanc2019cluster,
  title={A cluster-based operation model of aggregated battery swapping stations},
  author={{\v{S}}epetanc, Karlo and Pand{\v{z}}i{\'c}, Hrvoje},
  journal={IEEE transactions on power systems},
  volume={35},
  number={1},
  pages={249--260},
  year={2019},
  publisher={IEEE}
}

@article{hu2019collaborative,
  title={Collaborative optimization of distributed scheduling based on blockchain consensus mechanism considering battery-swap stations of electric vehicles},
  author={Hu, Wei and Yao, Wenhui and Hu, Yawei and Li, Huanhao},
  journal={IEEE access},
  volume={7},
  pages={137959--137967},
  year={2019},
  publisher={IEEE}
}

@article{wang2024optimal,
  title={Optimal participation of battery swapping stations in frequency regulation market considering uncertainty},
  author={Wang, Ziqi and Hou, Sizu},
  journal={Energy},
  volume={302},
  pages={131815},
  year={2024},
  publisher={Elsevier}
}

@article{wang2020vehicle,
  title={Vehicle to grid frequency regulation capacity optimal scheduling for battery swapping station using deep Q-network},
  author={Wang, Xinan and Wang, Jianhui and Liu, Jianzhe},
  journal={IEEE Transactions on Industrial Informatics},
  volume={17},
  number={2},
  pages={1342--1351},
  year={2020},
  publisher={IEEE}
}

@article{wang2020economic,
  title={Economic assessment for battery swapping station based frequency regulation service},
  author={Wang, Xinan and Wang, Jianhui},
  journal={IEEE Transactions on Industry Applications},
  volume={56},
  number={5},
  pages={5880--5889},
  year={2020},
  publisher={IEEE}
}

@article{wang2023short,
  title={Short-term electric vehicle battery swapping demand prediction: Deep learning methods},
  author={Wang, Shengyou and Chen, Anthony and Wang, Pinxi and Zhuge, Chengxiang},
  journal={Transportation Research Part D: Transport and Environment},
  volume={119},
  pages={103746},
  year={2023},
  publisher={Elsevier}
}

@article{du2025two,
  title={Two-layer decomposition-fused hybrid deep learning enables data-driven electricity demand forecasting for battery swapping station},
  author={Du, Pengcheng and Jiang, Meihui and Yang, Bowen and Chen, Baian and Zhu, Hongyu and Mengke, Qilao and Du, Yu and Kong, Fannie and Liu, Tianhao and Huang, Chao and others},
  journal={Energy},
  volume={332},
  pages={137288},
  year={2025},
  publisher={Elsevier}
}

@article{zhan2022review,
  title={A review of siting, sizing, optimal scheduling, and cost-benefit analysis for battery swapping stations},
  author={Zhan, Weipeng and Wang, Zhenpo and Zhang, Lei and Liu, Peng and Cui, Dingsong and Dorrell, David G},
  journal={Energy},
  volume={258},
  pages={124723},
  year={2022},
  publisher={Elsevier}
}

@article{amiri2018multi,
  title={Multi-objective optimum charging management of electric vehicles through battery swapping stations},
  author={Amiri, Saeed Salimi and Jadid, Shahram and Saboori, Hedayat},
  journal={Energy},
  volume={165},
  pages={549--562},
  year={2018},
  publisher={Elsevier}
}

@article{liu2021optimal,
  title={Optimal reserve management of electric vehicle aggregator: Discrete bilevel optimization model and exact algorithm},
  author={Liu, Wenjie and Chen, Shibo and Hou, Yunhe and Yang, Zaiyue},
  journal={IEEE Transactions on Smart Grid},
  volume={12},
  number={5},
  pages={4003--4015},
  year={2021},
  publisher={IEEE}
}

@article{he2025unlocking,
  title={Unlocking Ultra-Fast Charging: Reinforcement Learning with Reliability Guarantee for Ultra-Fast EV Charging Hub Under Behavior Uncertainty},
  author={He, Shangyang and Mai, Weijie and Tian, Jinpeng and Yang, Haosen and Liang, Zipeng and Wang, Chunhua and Li, Zhigang and Chung, Chi Yung},
  journal={IEEE Transactions on Transportation Electrification},
  year={2025},
  publisher={IEEE}
}

@inproceedings{fan2026ensuring,
  title={Ensuring Multi-Microgrids Emergency Resilience and User-Side Demand via Vehicle-to-X: A Transfer Hierarchical Safe Reinforcement Learning},
  author={Fan, Peixiao and Ma, Runzhuo and Bu, Siqi and Ke, Song and Wen, Yuxin},
  booktitle={2026 IEEE Green Technologies Conference (GreenTech)},
  pages={1--6},
  year={2026},
  organization={IEEE}
}

@article{mobarak2020solar,
  title={Solar-charged electric vehicles: A comprehensive analysis of grid, driver, and environmental benefits},
  author={Mobarak, Muhammad Hosnee and Kleiman, Rafael N and Bauman, Jennifer},
  journal={IEEE Transactions on Transportation Electrification},
  volume={7},
  number={2},
  pages={579--603},
  year={2020},
  publisher={IEEE}
}

@article{wang2023review,
  title={A review of electric vehicle auxiliary power modules: Challenges, topologies, and future trends},
  author={Wang, Cun and Zheng, Pengfei and Bauman, Jennifer},
  journal={IEEE Transactions on Power Electronics},
  volume={38},
  number={9},
  pages={11233--11244},
  year={2023},
  publisher={IEEE}
}

@article{gao2020deep,
  title={Deep reinforcement learning based optimal schedule for a battery swapping station considering uncertainties},
  author={Gao, Yuan and Yang, Jiajun and Yang, Ming and Li, Zhengshuo},
  journal={IEEE Transactions on Industry Applications},
  volume={56},
  number={5},
  pages={5775--5784},
  year={2020},
  publisher={IEEE}
}

@article{zipeng2017robust,
  title={Robust economic dispatch of microgrids containing electric vehicles},
  author={Zipeng, Liang and Haoyong, Chen and Yongchao, Wang and Cong, Zhang and Xiaodong, Zheng and Chulin, Wan},
  journal={Power System Technology},
  volume={41},
  number={8},
  pages={2647--2656},
  year={2017}
}

\end{document}